# Simulation of the 2018 Global Dust Storm on Mars Using the NASA Ames Mars GCM: A Multi-Tracer Approach

T. Bertrand[1], R. J. Wilson[1], M. A. Kahre[1], R. Urata[1], A. Kling[1]

[1]NASA/Ames Research Center, Moffett Field, California, USA

Corresponding author: Tanguy Bertrand (tanguy.bertrand@nasa.gov)

**Key Points:**

- We model the Mars Year 34 global dust storm with the NASA Ames Mars GCM in order to understand its onset, expansion and decay, and assess the sources and sinks of dust.
- The Hadley cell circulation and the diurnal cycle of atmospheric heating provide efficient dust transport upward and eastward, and both strongly increase in intensity as more dust is injected into the atmosphere.
- We highlight the back and forth transfer of dust between reservoirs located in the Arabia/Sabaea and Tharsis regions, which may play an important role in the development of the storm by replenishing the surface dust reservoirs.
- We highlight a strong sensitivity of the positive-radiative feedbacks between dust lifting, atmospheric heating and the strengthening circulation to the dust particle size.
- We find that the upper atmosphere is enriched in water vapor as a result of thinner water ice clouds that migrated to higher altitudes as the atmosphere warms in response to dust heating.

**Abstract**

Global dust storms are the most thermodynamically significant dust events on Mars. They are produced from the combination of multiple local and regional lifting events and maintained by positive radiative-dynamic feedbacks. The most recent of these events, which began in June 2018, was monitored by several spacecraft in orbit and on the surface, but many questions remain regarding its onset, expansion and decay. We model the 2018 global dust storm with the NASA Ames Mars Global Climate Model to better understand the evolution of the storm and how the general circulation and finite surface dust reservoirs impact it. The global dust storm is characterized by the rapid eastward transport of dust in the equatorial regions and subsequent lifting. We highlight the rapid transfer of dust between western and eastern hemispheres reservoirs, which may play an important role in the storm development through the replenishment of surface dust. Both the Hadley cell circulation and the diurnal cycle of atmospheric heating increase in intensity with increasing dustiness. Large dust plumes are predicted during the mature stage of the storm, injecting dust up to 80 km. The water ice cloud condensation level migrates to higher altitudes, leading to the enrichment of water vapor in the upper atmosphere. In our simulations, the intensity of the Hadley cell is significantly stronger than that of non-dusty conditions. This feedback is strongly sensitive to the radiative properties of dust, which depends on the effective size of the lifted dust distribution.

**1 Introduction**

Global dust storms (GDS) are the largest spatial-scale dust lifting events on Mars and represent one of the most puzzling phenomena of the Mars dust cycle. They occur only every few Martian years (MY), during Mars' dusty season (i.e., northern fall and winter; solar longitudes Ls=185°-300°) and usually mask most of the surface from orbit for several months as they inject large amounts of dust into the atmosphere and produce high dust visible optical depths (typically larger than 3 and up to 10). Global storms appear to have a distinct climatology from the more regular seasonal cycle of pre- and post-solstice regional storms, the so-called A and C season dust activity by *Kass et al*. [2016].

Following the start of global mapping in Mars Year 24 (1999) by Mars Global Surveyor (MGS), Mars has been under near-constant surveillance by one or more spacecraft up to the current MY35, providing detail on the impact of dust storm activity on the Martian climate. Dust storms and the subsequent atmospheric dust loading significantly warm the atmosphere of Mars and thus alter the atmospheric circulation as well as the $CO_2$ and water cycles [e.g. *Wilson*, 1997, *Newman et al*., 2002, *Guzewich et al.*, 2014, *Strausberg et al*., 2005, *Kahre et al.*, 2017]. In particular, global dust storms (MY25, MY28, MY34) have an impressive effect on atmospheric temperatures. Figure 1 summarizes this impact on equatorial zonal-mean $T_{15}$ temperatures, representative of a depth-weighted layer of atmosphere centered at ~30 Pa (see details on $T_{15}$ in Section 2.1). Such global storms also have a global impact on surface properties. For instance, changes in surface albedo have been observed after the occurrence of GDSs [e.g., *Cantor*, 2007, *Szwast et al*., 2006, *Vincendon et al*., 2015], suggesting a redistribution of the surface dust reservoirs (dust deposits tend to brighten the surface). These transfers of dust from different surface reservoirs are key as they may create the hysteresis responsible for the inter-annual variability of GDSs [*Kahre et al*., 2005, *Newman et al*., 2015].

The most recent global dust storm, the 2018/MY34 GDS, was observed by several instruments, including the Mars Color Imager (MARCI) on-board Mars Reconnaissance Orbiter (MRO), which produced daily global maps of Mars during this period [*Malin,* 2018, a-f], and by the Mars Climate Sounder (MCS), which produced profiles of temperature and aerosol [*see paper in this special issue*]. This GDS is characterized by a relatively early season of onset (near the northern fall equinox around $L_s=180°$), and much larger amplitudes in opacity and temperature (although not warmer than MY25) than previous GDSs (Figure 1). In fact, GDSs were long viewed as solstitial season phenomena. Modeling studies of GDSs have generally focused on simulating storms in the solstice season (onset around $L_s=270°$) due to the expectation that the Hadley circulation, which is most intense in this season, triggers efficient lifting, transport of dust and associated positive feedbacks and thus drives the genesis of the GDSs [e.g. *Haberle*, 1982, *Basu et al.*, 2006, *Kahre et al.*, 2005, *Newman and Richardson*, 2015]. However, the 2001/MY25 and 2018/MY34 events call for a change in that view. The 2007/MY28 storm is the only solstitial storm seen in the era of continuous spacecraft monitoring (MY23-MY34).

Although the 2018/MY34 GDS was monitored by several instruments in orbit and on the surface of Mars, many fundamental questions remain unsolved regarding its onset, evolution, and impact on the Martian climate. In particular: What controls and triggers the onset of the GDS? How does the GDS expand, and why and by which mechanisms does it stop expanding? Where are the surface dust reservoirs and how is dust lifted and transferred from one reservoir to another? What are the differences between the MY34 and MY25 GDSs events, and how are these two storms distinguished from the more regular regional storms that develop in the A-season ($L_s\sim210°$-$235°$)? How rapidly are finite dust reservoirs depleted and restored? How are lifting centers remotely triggered? These scientific questions call for modeling efforts of the MY34 GDS.

In this paper, we aim to provide new insights on these questions and more generally on the evolution of the present-day dust cycle on Mars by simulating the MY34 GDS with the NASA Ames Mars Global Climate Model (MGCM). MGCMs are valuable because we can relate temperature observations to circulation elements and subsequently investigate feedbacks between dust lifting, surface stresses and aerosol transport. Here our goals are to capture the storm evolution, qualitatively and quantitatively assess the sources, sinks and transport of dust, study the sensitivity of the circulation to different parameters, and further our understanding of the dust cycle on Mars. In addition, the MY34 event represents a test bed for assessing new model capabilities such as the tagging method, which allows dust to be partitioned in a variety of ways to assess aspects of geographical and temporal sources and physical lifting processes (Section 3). While we investigate the storm using this method, this effort remains a work in progress.

We first describe the observational metrics used in this paper (MCS brightness temperatures and opacity maps) and give an overview of the MY34 global dust storm in Section 2. We then present the model used and describe the settings of the reference simulation in Section 3. Section 4 gives an overview of the MY34 GDS as simulated by our reference simulation, compares the results with the available observations and discusses the discrepancies. Section 5 provides an analysis of the dust sources, sinks, and transport obtained in this simulation. Finally, in Section 6, we discuss our results further in light of a sensitivity study and comparisons with other regional and global dust storms.

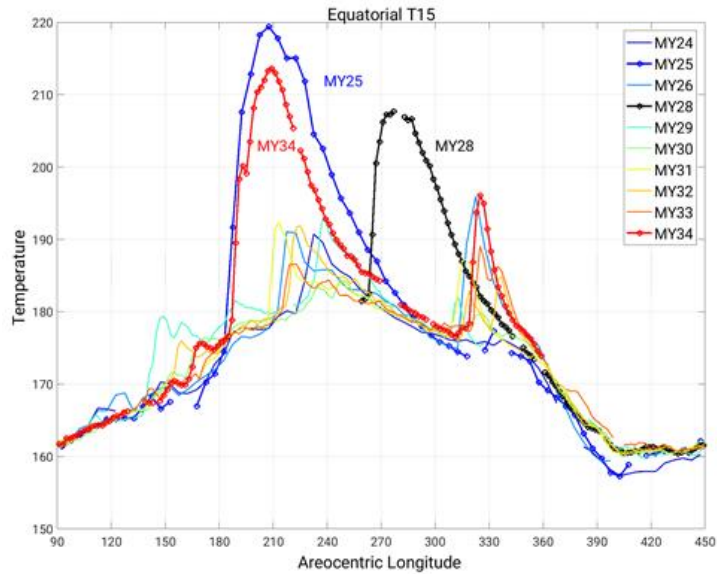

Figure 1: The seasonal variation in zonally averaged equatorial $T_{15}$ from 3 pm MCS limb observations for Mars Years 28 to 34. These MCS temperatures are based on 15 micron brightness temperatures in limb viewing mode, representing a depth-weighted atmospheric temperature at ~30 Pa. Also shown are TES temperatures at 30 Pa for Mars Years 24 to 26. The MY34 storm is a significant outlier in the normal cycle of A and C season regional storm activity.

## 2 Background: Observations of the MY34 Global Dust Storm

In this section we give a brief overview of how the 2018/MY34 GDS evolved based on the available observations (MARCI, MCS). Other descriptions of the storm can be found in *Guzewich et al.* [2019], *Sanchez-Lavega et al.* [2019], and in further detail in other papers of this special issue.

2.1 Description of the Set of Observations

*2.2.1 Dust Scenarios*

Spacecraft observations have enabled the creation of a daily record of spatially-resolved column dust IR opacity from mostly TES and MCS observations for the last 10 Martian years (Montabone et al. 2015). In the case of the MGS mission, these fields are primarily derived from TES nadir observations of column opacity. For MRO, the column opacity is primarily derived from the downward extrapolation of profiles of dust opacity retrieved from limb viewing observations.

These sets of gridded, daily dust opacity maps are an important input for MGCM simulations (see Section 3) and will be referred to in this paper as dust scenarios. Here we use the most recent dust scenario for MY34, derived from the available opacity observations provided by TES, MCS, The Thermal Emission Imaging System (THEMIS, onboard Mars Odyssey), and lander observations. The version of this scenario is V3-2_beta [*Kleinboehl et al.,* 2017,

*Montabone et al.*, 2019]. We have converted the IR (9 µm) opacities to visible opacities for comparison with visible opacities typically reported by MGCM modeling groups.

### 2.2.2 $T_{15}$ Temperatures

In this paper, in order to illustrate the impact of dust opacity on global scale atmospheric temperature, we compare simulated and observed brightness temperatures. The observations are derived from MCS radiances from the lowest detector in the 15 µm (A3) channel included in the level 1 data product in the Planetary Data System archive. We refer to this measure as $T_{15}$, which represents a depth-weighted temperature centered at 40 Pa (~30 km) where the A3 channel weighting function peaks. The on-planet vertical weighting function is similar to that of the Viking IRTM 15 µm channel [*Wilson and Richardson*, 2000]. The use of brightness temperature bypasses the temperature retrieval process required for profiling and yields results that are uninfluenced by aerosols. Thus this measurement allows relatively low altitude nighttime and daytime temperatures (MCS observations are obtained from a sun-synchronous orbit, providing observations at 3 am and 3 pm mean solar time) to be obtained for all latitudes throughout the MRO mission. This is particularly useful for examining the temperature response to dust storm events, as the diurnal temperature range and decay of average temperature is diagnostic of the influence of dust heating [*Conrath*, 1975].

### 2.2 Overview of the MY34 GDS

The 2018/MY34 GDS lasted about 110 sols, starting mid-May ($L_s$ ~181°, northern spring equinox) and ending around mid-September ($L_s$ ~ 250°). As observed during previous GDS events (MY25, MY28), the MY34 GDS was not composed of a single lifting event but rather of several local and regional storms.

MARCI observations revealed several local frontal-like dust storms occurring along the seasonal cap edge in the high-latitude plains of the northern hemisphere around $L_s$ ~ 181°. These frontal storms, driven by baroclinic activity [*Wang and Richardson*, 2015, *Wang et al.,* 2003], are likely to be the precursors of the MY34 global dust storm. At this season, the general circulation is characterized by a two-cell Hadley circulation with a narrow ascending branch near the equator and broader descending branches at high latitudes. The descending branch in the high-northern latitudes confined the frontal dust storms to the lowest atmospheric levels, and transported the lifted dust towards southern latitudes. Interestingly, transient waves in the southern hemisphere likely contributed to the enhancement of dust in Hellas Basin ($L_s$=179°-183°), with subsequent triggering of lifting in Tyrhenna [*Straussberg et al.*, 2005].

Around $L_s$ ~ 185°, the frontal storms penetrated the tropics in Acidalia/Chryse Planitia (30°N–60°N, 300–360°E), one of the two low-topographic "flushing" channels known to be efficient at transporting dust into the south hemisphere (Utopia/Isidis is the second one) and to be an active region for dust lifting and onset of regional storms [*Wang and Richardson,* 2015]. Note that this is the first monitored GDS observed to initiate in the northern hemisphere (MY25 and MY28 initiated in the southern hemisphere, in Hesperia Planum and Chryse/Noachis respectively). Around $L_s$ ~ 187°, the accumulation of atmospheric dust in Acidalia/Chryse Planitia formed a regional storm, responsible for the sudden and abrupt increase in the observed global mean dust opacity and atmospheric temperatures (Figure 1).

From there, the storm expanded eastward and southward. Distinct dust storms also occurred along the receding southern seasonal $CO_2$ polar ice cap [e.g., *Toigo et al.*, 2002], in particular near the region south of Hellas ($L_s$=188°-192°), as revealed by MARCI observations [*Cantor et al.,* 2019, *Malin et al.*, 2018a-g] and to some extent by the MCS-derived opacity maps. By $L_s$=192°, these storms merged with the larger regional tropical storm along and north of the equator, which then expanded further east and became global by $L_s$=193°. Note that this eastward expansion in the tropics is a dominant feature of the MY25 and MY34 dust storms, and is not observed in the A and C season regional storms [*Wang and Richardson*, 2015]. From the Hadley cell circulation point of view, the conservation of angular momentum should lead to strong extratropical prograde (eastward) jets and a weak retrograde (westward) equatorial jet in the upper atmosphere (if ones assumes no friction with the surface before the air rises up, which would lead to perfect conservation of angular momentum). However, in the tropics, diurnal thermal tides tend to force an eastward zonal-mean flow, stronger above the equator at about 10-20 km above surface [*Wilson and Hamilton*, 1996, *Lewis and Read*, 2003]. This flow intensifies with stronger dust forcing such as that induced by the onset of the GDS.

As dust lifting continued, dust visible opacities locally peaked between 5 and 10. The global mean dust opacity reached a peak of ~4 around $L_s$=205°-210°, which corresponds to the maximum thermal impact of the storm, based on tide analysis and observed MCS $T_{15}$ temperatures and derived column opacities. From $L_s$=210°, the GDS entered the decay phase and dust started settling out of the atmosphere. Atmospheric dust concentrations returned to nominal seasonal levels by $L_s$ ~ 250°.

**3 Model Description**

3.1 The NASA Ames Mars GCM

We use the NASA Ames Mars GCM (MGCM), which now employs (1) the NOAA/GFDL cubed-sphere finite-volume (FV3) dynamical core and (2) physics packages from the Ames Legacy MGCM as described in *Kahre et al*. [2018] and *Haberle et al*. [2019]. The cubed-sphere grid is relatively uniform, which enables efficient high-resolution simulations on massively parallel computers. The model uses topography from the Mars Orbiter Laser Altimeter (MOLA) and albedo and thermal inertia maps derived from Viking and Mars Global Surveyor (MGS) Thermal Emission Spectrometer (TES) observations.

The model includes fully coupled dust and water cycles, using the Ames water ice cloud microphysics package described in *Haberle et al*. [2019]. This includes water sublimation from the north polar residual water ice cap and the complex processes of cloud microphysics (nucleation, growth, settling) [*Montmessin et al.*, 2002, 2004, *Nelli et al.,* 2009, *Navarro et al.,* 2014]. The airborne dust that interacts with solar and infrared radiation acts as ice nuclei and goes through gravitational sedimentation as free dust and as cores of water ice cloud particles. The lognormal particle size distributions of dust and clouds are represented by a spatially and temporally varying mass and number, and a constant effective variance (two-moments scheme). Many different dust lifting schemes are implemented, based on observations or equations representing the processes of convective (dust devils) and wind stress lifting ("interactive dust lifting", e.g, *Kahre et al.* [2006, 2015]). Here we use the available dust scenarios (observed column opacity fields) as a constraint for the simulation to match ["assimilated dust lifting", e.g. *Kahre et al.*, 2009, *Greybush et al.*, 2012].

The planetary boundary layer (PBL) model in the MGCM solves energy and momentum equations to predict wind and temperature profiles in the atmosphere and mix tracers. It employs a Mellor-Yamada level-2 boundary layer scheme for turbulence closure. This implementation is first described in *Haberle et al.* [1993], and later updated in *Haberle et al.* [1999]. The model calculates surface fluxes based on heat and momentum drag coefficients from the stability functions from *Savijarvi* [1995], and *Hourdin et al.* [1995]. Eddy mixing coefficients are calculated using the equations from *Arya* [1988].

The model employs a 2-stream radiative transfer scheme that accounts for gaseous absorption of $CO_2$, $H_2O$ and scattering aerosols, including dust and water ice particles [*Toon et al.*, 1989, *Haberle et al.*, 2019]. Gaseous opacities are calculated using the correlated-k method, and Rayleigh scattering is calculated for $CO_2$. Mie theory is used to calculate extinction efficiencies and scattering properties for aerosols, assuming a log-normal size distribution. Ice is assumed to have a core-mantle structure, with refractive indices for dust taken from *Wolff et al.* [2010], and ice from *Warren* [1984]. We note that the optical properties of both dust and water ice cloud particles depend on particle size and thus evolve in time and space. In the case of dust, the assumed lifted dust particle size distribution affects key aspects of the radiative behavior of the airborne dust grains (e.g., the visible to infrared ratio). This issue is discussed further in Section 4.1. The optical properties are used in a two-stream code to calculate fluxes, and heating rates are computed from the flux divergences. There are 7 visible bands (0.4-4.5 μm) and 5 IR bands (4.5-1000 μm).

## 3.2 The Tagging Method

The MGCM includes a numerical tagging method [*Bertrand et al.*, 2018], which "tags" (or labels, or follows) any atmospheric constituent (the "reference tracer", e.g. dust particles, water vapor, water ice, argon etc.) according to a chosen criterion (e.g. location, local time, type of lifting, amplitude of the dust source or wind stress, reached altitude, etc.). Each tag is transported by the circulation and behaves as the constituent it follows but remains completely passive and does not alter the predictions. This technique enables us to track not only the origin of a given atmospheric constituent, but also the physical processes it goes through (e.g., scavenging, formation of ice clouds, frost, etc.), or the different environments it has encountered since its emission (craters, mountains, dusty atmosphere, poles, etc). This powerful method was first implemented in the NASA/GISS GCM [*Koster et al.*, 1986] to identify the origin of the precipitation in various regions of the Earth, and is now widely used for detailed studies of the Earth's water cycle. It has never before been tested on Mars, and has plenty of promising applications.

## 3.3 Reference Simulation Settings

### *3.3.1 Overview*

The simulations described in this paper were carried out with a horizontal resolution of 2°x2° and 46 vertical levels, with a vertical resolution decreasing from 20 m near the surface to 10 km at the model top (~80 km). The reference simulation is performed with two-moment aerosol sizes distributions for water ice and dust. Dust particles are represented by a lognormal particle size distribution with an effective variance of 0.5 and an effective particle radius of 3 μm [*Clancy et al.*, 2003, *Wolff and Clancy,* 2003, *Kahre et al.,* 2015]. Here, the effective particle radius is larger than the 2.5 μm value previously used in *Kahre et al.* [2015]. This choice is driven by better agreement between simulated and observed opacities, as highlighted in the sensitivity study detailed in Section 6. The simulation has been carried out over multiple annual cycles (using prescribed dust map scenarios for MY33) so that the aerosol and temperature distributions could reach a seasonally equilibrated state before running the simulation for MY34. We use radiatively active dust, water vapor and water ice clouds.

### 3.3.2 Dust Lifting Scheme

There are several possible strategies for dust lifting used for controlling and influencing the 3-D distribution of aerosol in a MGCM simulation. Here we use the maps from the dust scenario, interpolated in time and on the GCM grid, to allow for the GCM to identify dust lifting centers and reproduce realistic opacities and temperatures. Dust is injected from the surface (we assume an infinite reservoir across the globe) into the PBL when the simulated dust column opacity is lower than that in the dust scenario [*Kahre and Wilson*, 2009, *Greybush et al*., 2012, *Bertrand et al*. 2019] and is then allowed to be transported elsewhere by the simulated general circulation. The amplitudes of the dust sources are calculated so that the model tracks the observed dust column opacities. We allow dust to be lifted at night, but not over the polar caps. Note that in this reference simulation, dust is only removed from the atmosphere back onto the surface by sedimentation processes. We do not remove dust artificially (from the PBL or by rescaling the vertical distribution) if the simulated opacity exceeds the prescribed opacity from the dust scenario. This is because we want to realistically represent the processes of dust lifting and the pathways of dust transport, and not include physics we do not understand. Using artificial sinks of dust or other types of data assimilation does not provide proper insight into the evolution of thermal tides, the vertical dust distribution, or surface stress.

### 3.3.3 Dust Tagging

In order to better understand the pathways of dust and better characterize the transfer of surface dust between different reservoirs during the GDS, we used the tagging method to track dust in the reference simulation according to its location of lifting. Figure 2 shows the different regions that we consider in this paper. These regions correspond to the most active dust lifting centers during the MY34 GDS. We also tagged dust injected into the atmosphere from the intense lifting occurring in the Tharsis regions around $L_s$=198°. This allows us to follow how and where dust is transported in the atmosphere during the mature stage of the GDS.

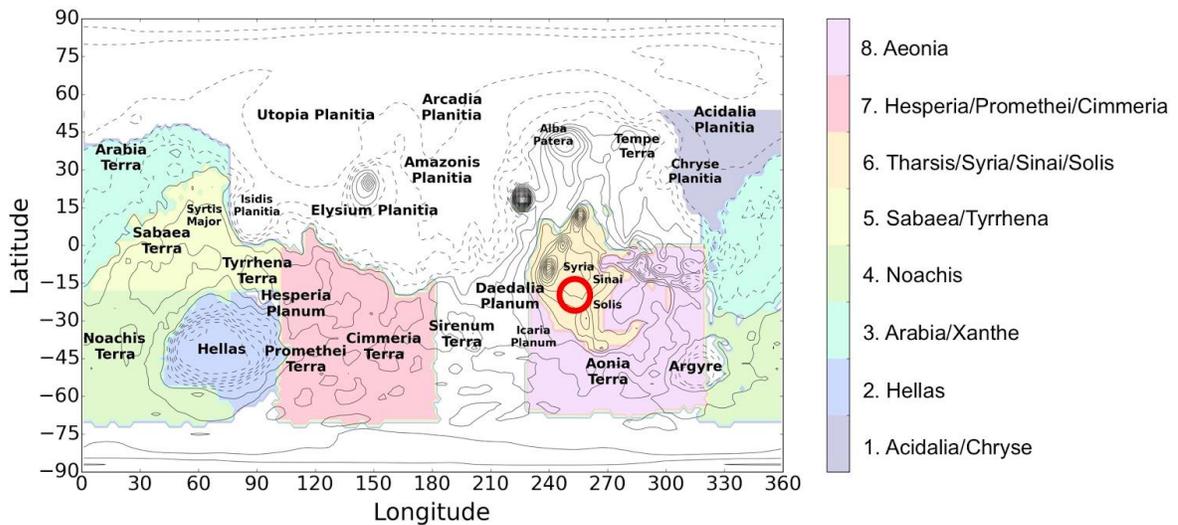

Figure 2: Map of Mars showing the main regions of surface dust lifting during the MY34 GDS. We use this map to track dust that has been lifted from each of these regions. We also track one plume of dust (see Section 5.3) that originates from intense dust lifting near the Tharsis region around $L_s$=198°, as indicated by the red circle.

**4 Model Results: Validation and Overview of the MY34 GDS Phases**

In this section, we compare opacities and temperatures predicted by the GCM with the MCS observations, discuss the discrepancies, and then present an overview of our best-case reference GCM simulation of the MY34 GDS.

4.1 Validation: Opacities and Temperatures in the GCM vs Observations

Figure 3 compares the zonal mean observed column dust visible opacities (from the dust scenario, see Section 2.1) with those obtained with our reference GCM simulation. The simulated opacities are in reasonable agreement with the observations, and capture the abrupt increase of opacity around $L_s$=187°-188° and a peak of opacity around $L_s$=210°. However, simulated opacities are slightly lower than the observations during the onset of the storm and larger during the decay phase of the storm (with opacity differences up to 0.2).

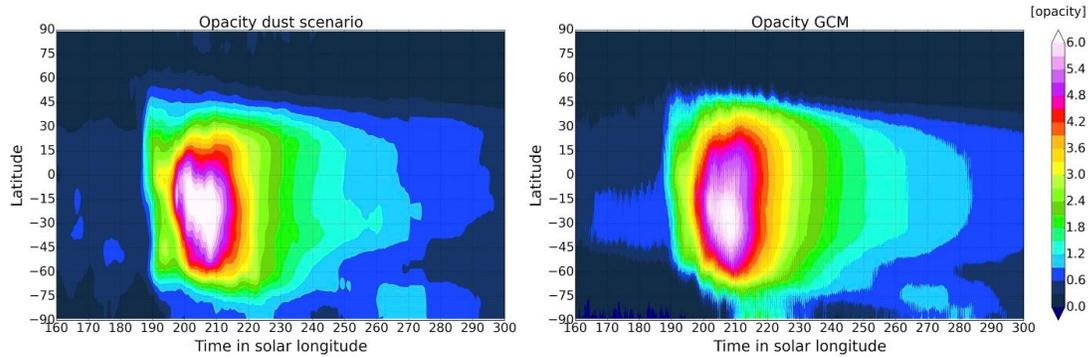

Figure 3: Zonal mean column dust visible opacities from the MCS-derived dust scenario (left) and the reference GCM simulation (right) during the MY34 GDS.

This is also shown in Figure 4, which compares observed and simulated maps of column dust visible opacity during the different phases of the GDS. The model captures the main regional storm in Acidalia/Chryse/Arabia during storm onset, the large opacities in the Tharsis/Aonia regions during the mature stage of the GDS, and comparable opacities with observations at all latitudes. However, simulated opacities around $L_s$=189°-195° locally peak at ~4 while the observed opacities peak at ~5 during the same period and at the same locations. At $L_s$=207°, simulated globally averaged opacities are larger than those observed, but locally peak at ~9 while the observed opacities peak at ~10 in the Tharsis region and Sabaea Terra. These discrepancies are reflected in the temperatures, as shown in Figure 5. Simulated daytime and nighttime $T_{15}$ temperatures are in good agreement with the observed MCS $T_{15}$ temperatures during the onset of the storm but are ~10 K warmer during the peak activity of the storm around $L_s$=201°-207° and during the decay phase of the storm.

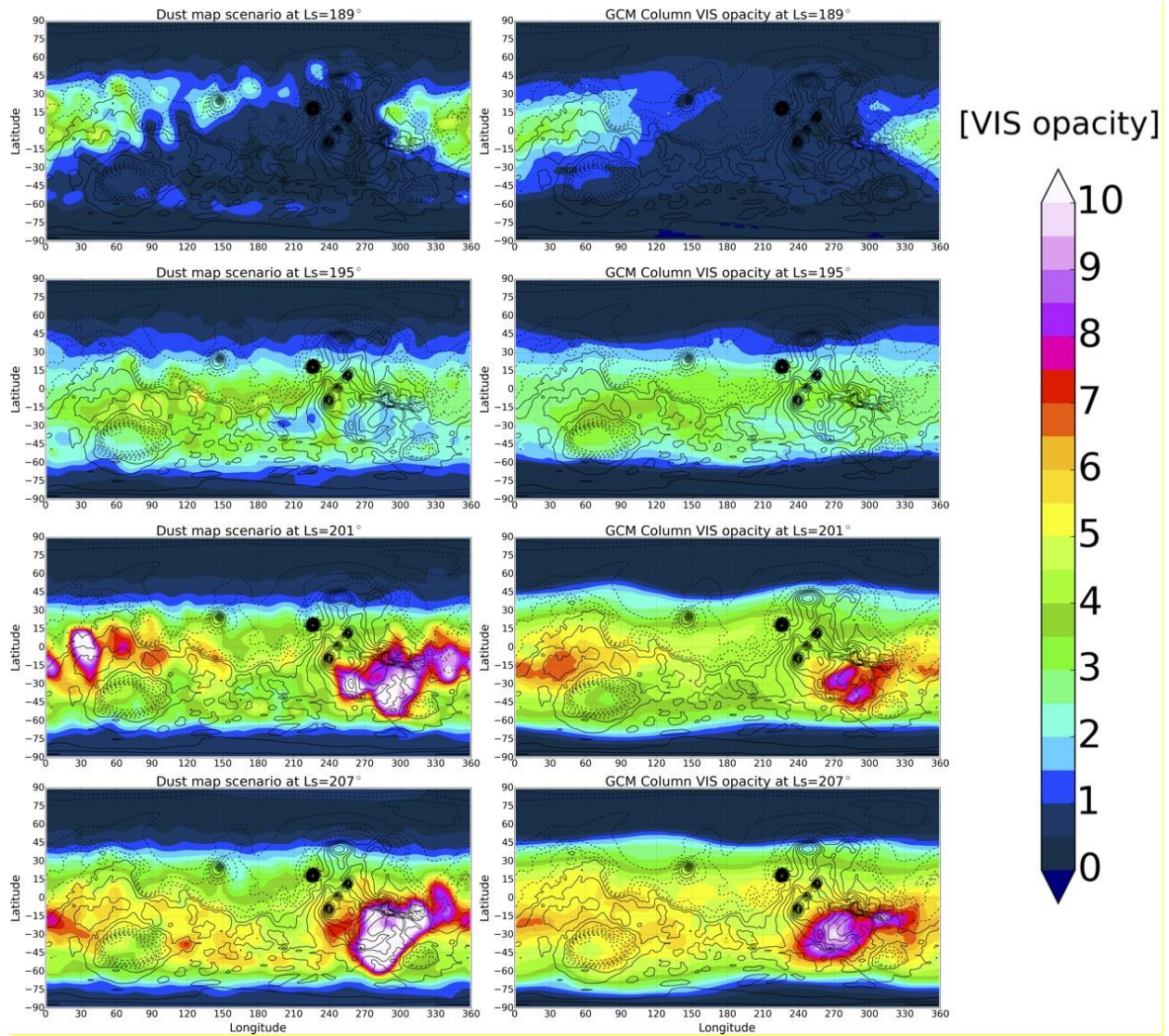

Figure 4: Maps of column dust visible opacities from the MCS-derived dust scenario (left) and the reference GCM simulation (diurnal mean, right) during different phases of the MY34 GDS: onset ($L_s=189°$), global expansion ($L_s=195°$), maximum dust lifting south of Tharsis ($L_s=201°$), mature stage and peak global dust opacity ($L_s=207°$).

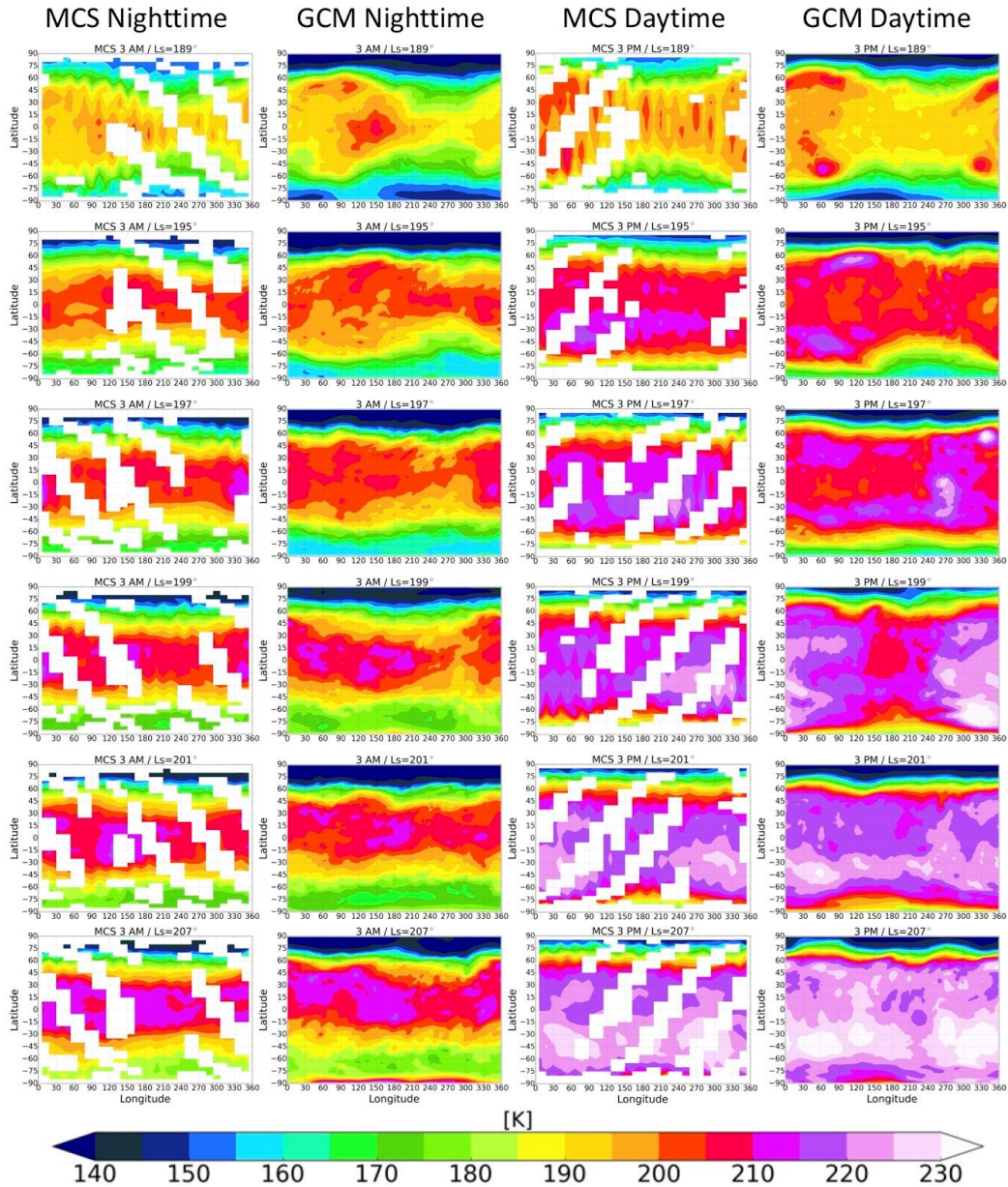

Figure 5: Maps of 3am (left two columns) and 3 pm (right two columns) brightness temperatures from MCS observations (columns one and three) and the reference GCM simulation (columns two and four) at solar longitudes $L_s$ = 188°, 193°, 197°, 199°, 201°, and 207°.

4.2 Possible Reasons for Discrepancies Between Model Predictions and Observations

The differences between the model-predicted and observed column dust opacities during the onset of the GDS (i.e., slightly lower in the simulations than in the dust scenario) are due in part to an overly large lifted dust effective radius (3 µm) used in the GCM. Larger particles settle rapidly to the surface shortly after being lifted from the surface, which results in a slower increase in atmospheric dust than what is needed to match the prescribed dust opacity. Using lifted dust distributions with smaller particles help resolve this issue. However, simulations with smaller lifted dust effective radii result in a much longer decay phase of the GDS (as smaller particles settle out more slowly), which is less consistent with observations. This is illustrated by Figure 6a. In our reference simulation, the selected effective radius of lifted dust distribution (3 µm) is therefore the best choice to produce (1) reasonably strong lifting during the onset and development of the GDS, (2) realistic timescales for the sedimentation of the airborne dust during the decay phase of the GDS, and (3) atmospheric $T_{15}$ temperatures in agreement with MCS observations. This choice has consequences on the strength of the Hadley cell circulation, as discussed in Section 4.3 and 6.1.

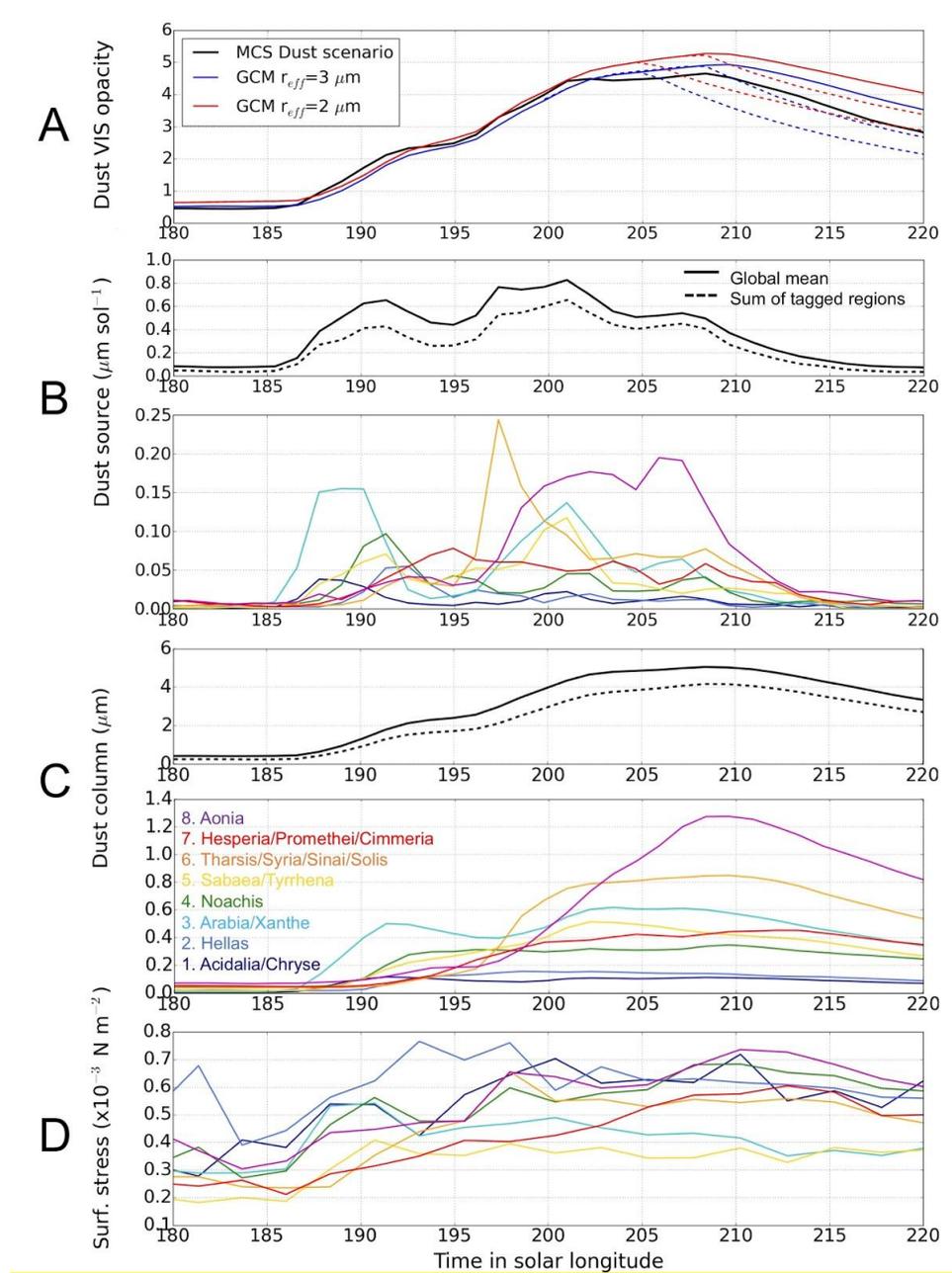

Figure 6: A. Global mean column dust visible opacity during the MY 34 GDS from the MCS observations (dust scenario, black line), the reference GCM simulation (solid blue), and several GCM sensitivity simulations. The solid red line corresponds to a simulation with a lifted dust particle effective radius of 2 microns. The dashed lines show the evolution of opacity if all dust sources are shut down in the simulations after $L_s=205°$ or $L_s$ 208° assuming a lifted dust particle effective radius of 2 microns (red) or 3 microns (blue). In these cases, the global opacity exhibits an exponential decay that brackets the observed opacity during the decay phase of the storm. B. Global mean dust lifting rate (i.e., sources; top) and the contribution from each tagged region (bottom). C. Global mean dust atmospheric column mass (top) and the contribution from each tagged region (bottom). D. Normalized maximum surface stress over 2 sols in each tagged region. The color scales are the same as those used in Figure 2.

In reality, it is possible that small particles are lifted during the GDS onset, and then larger particles are progressively lifted as the storm develops and surface wind stresses increase. This could explain why the GCM simulation underestimates the column dust opacity during the onset of the storm and overestimates it during the decay phase. Alternatively, some surface dust reservoirs could contain dust particles of different sizes, thus significantly changing the dust size distribution in the atmosphere over the course of the GDS as dust lifting becomes active in these reservoirs. In the future, using a multi-modal dust distribution and adapting the lifting schemes could be of interest to investigate these processes and improve the simulations.

During the GDS, the model generally predicts lower dust opacities than observed where high opacities are prescribed by the dust scenario and higher opacities where low opacities are prescribed. These discrepancies could be due to insufficient vertical mixing by the GCM or erroneous opacities in the dust scenario (see below). For instance, near the southern polar cap edge, local dust lifting in the real atmosphere is presumably caused by strong near-surface winds associated with large temperature gradients. In the GCM, the general circulation continuously transports dust away from the polar cap edge, making it impossible to locally match the prescribed opacities. Implied cap edge dust lifting may therefore be overestimated in our simulations, which leads to an overestimate of column dust opacities elsewhere, as the regions near the cap edge continuously feed the atmosphere with an excess of dust. Similarly, during the mature phase and the decay phase of the MY34 GDS, dust lifting could be overestimated in some regions (e.g. Tharsis, see Section 5), leading to a higher opacity than that observed by MCS.

The discrepancies between the observed and simulated opacities and temperatures could also be explained in part by the fact that the MCS-derived dust scenario may contain imperfections. First, the opacity maps are relatively patchy, which is likely unphysical and relates to an observational bias. As the GCM simulation tries to reproduce these opacities, it leads to patchy dust lifting sources and local underpredicted and overpredict opacities. Second, some opacities may be overstated. In particular, the amount of dust lifted in the GCM is very large and may be unrealistic in the Tharsis region around $L_s=198°$. (See Section 5.3). This could be due to unrealistic prescribed opacities at this time and location. However, we note that the simulated atmospheric brightness temperatures during this period are in agreement with observations (Figure 5), which show 10-30 K higher temperatures above the Tharsis region during $L_s=197°$-199° (during daytime), thus suggesting that the amounts of dust predicted by the GCM at this time and over this region are not overestimated. Third, the opacity maps do not capture the frontal storms that are seen in MARCI images in the high northern latitudes before the onset of the GDS. Other storms occurring near the southern polar cap edge during the GDS may be missing from the opacity maps. This is most likely because these storms are local and rapidly evolving events (the opacity maps correspond to diurnal averages gridded to 4° resolution). Additionally, they are shallow dust storms, confined to the lowest atmospheric levels by the descending branch in the high-northern latitudes. These two characteristics make them challenging to detect with MCS limb observations. However, we note that frontal storms may also be too short and too shallow to significantly impact the dust cycle, atmospheric temperatures and the general circulation, so this may not be a key issue for our GCM simulation.

4.3 Sensitivity Studies

We performed several simulations to test the sensitivity of the GCM results to several parameters and modeling approaches in order to try to solve the issues listed above. Changing the lifted dust particle effective radius (in steps from 0.5 µm to 5 µm) and the effective variance of the dust particle population (0.3-0.7) improves either the onset or the decay phase of the storm, but not both at the same time. For instance, injecting smaller particles into the atmosphere allows for better agreement between model results and observations of opacities during the onset of the storm, but leads to a longer decay and larger discrepancies during this period. Shutting down the dust lifting in the GCM during the mature stage of the GDS results in a better agreement between simulated and observed opacities. Figure 6a shows a selection of simulations that bracket the observations. We note that changing the lifted dust particle effective radius impacts the strength of the Hadley cell, which is discussed in Section 6.1.

Injecting dust instantaneously to higher altitudes (up to 25 km) than the computed PBL height slightly reduces the discrepancies during onset, but slightly increases them during the mature stage of the storm. Removing the highest opacity peaks in the prescribed column dust opacity maps (e.g. limiting the maximum column dust visible opacity to 6, 8 or 10) leads to better agreements in opacities and temperatures during the mature stage and the decay phase of the GDS, but does not solve the issues during onset and expansion (as lower opacities are prescribed). Allowing dust injection during daytime only or during specific ranges of local times does not significantly change the results. The same applies if we limit cap edge lifting, or if we allow dust injection over the polar $CO_2$ ice caps to match the prescribed column dust opacities in those regions. Tuning the intensity of the dust injection from the surface (by a factor ranging from 0.5 to 1.2) improves the agreement between the observed and simulated global mean opacities but locally significant discrepancies are still obtained. The amount of water vapor and ice clouds and their radiative impact also have little effect on the global evolution of the simulated storm. We also tried a simulation using a 1°x1° horizontal resolution, which did not change the results, mostly because the maps of the dust scenario are interpolated to a 3.75°x3.75° grid [*Montabone et al.*, 2015].

In summary, our results remain in a generally good agreement with the observations although we acknowledge small and local discrepancies between the observed and simulated opacities and temperatures. In particular, it has been difficult to precisely match the global mean opacities from the dust scenario at the beginning (onset) and the end (decay) of the global dust storm. Parts of these issues are attributed to erroneous opacities in the dust scenario and to modeling approximations (e.g., related to the way dust is lifted from the surface or vertically transported in the GCM). Solving these issues may require bi-modal and multi-modal dust distributions that can be seen by the radiative transfer. However, all of our GCM simulations described in this section predicted a similar evolution of the GDS and overall, the pathways of dust lifting and the pattern for transport, sources and sinks of dust remain unchanged. This demonstrates the robustness of our GCM results.

4.4 Overview of the Different Phases of the MY34 GDS

Figure 7 shows the variation of the zonally averaged dust lifting rate with time as predicted by our reference simulation. The GDS is characterized by zonally averaged dust lifting rates of greater than 0.3 µm per sol, occurring from $L_s=187°$ to $L_s=210°$ (before the decay phase) and from latitudes 60°S to 60°N. We note that the storm defined in this way appears to form an "ape" shape on Figure 7, which we use to divide the GDS in four main phases: (1) The onset of the storm (corresponding to the forearm of the ape), occurring between latitudes 5°S and 60°N and from $L_s=187°$ to $L_s=193°$. This phase corresponds to a regional dust storm that develops and moves southwards in Acidalia/Chryse Planitia and Xanthe Terra. Note that significant lifting along the edge of the retreating south polar cap also occurs during this period. (2) A rapid eastward and southward expansion of the storm (corresponding to the shoulder of the ape), occurring below the equator from $L_s=193°$ to $L_s=196°$. During this period, the regional storm turns into a global storm, triggering dust lifting at all longitudes. (3) A period of intense and maximum dust lifting (corresponding to the head of the ape), occurring from $L_s=196°$ to $L_s=204°$ between latitudes 20°S and 15°N in the Tharsis/Thaumasia plateau region. Intense dust lifting during this period is also predicted around 60°S, in the Sabaea/Tyrrhena region and in Aonia Terra. (4) The mature stage of the storm, where maximum atmospheric dust loading (shoulder of the ape) and maximum atmospheric temperatures occur. Most of the lifting during this period occurs in Aonia Terra and the Tharsis region.

During the pre-storm period, dust lifting rates are obtained within the range 0.01-0.1 µm per sol at all latitudes, except near the southern polar cap where the dust lifting rate peak at about 1 µm per sol. During the decay phase of the storm, little to no dust is lifted in the tropics. This is related to the main issue of the simulation, discussed in Section 4.2 and 4.3: the model is unable to reproduce a reasonable decrease in dust opacity during the decay phase of the GDS and tends to predict opacities significantly larger than those observed.

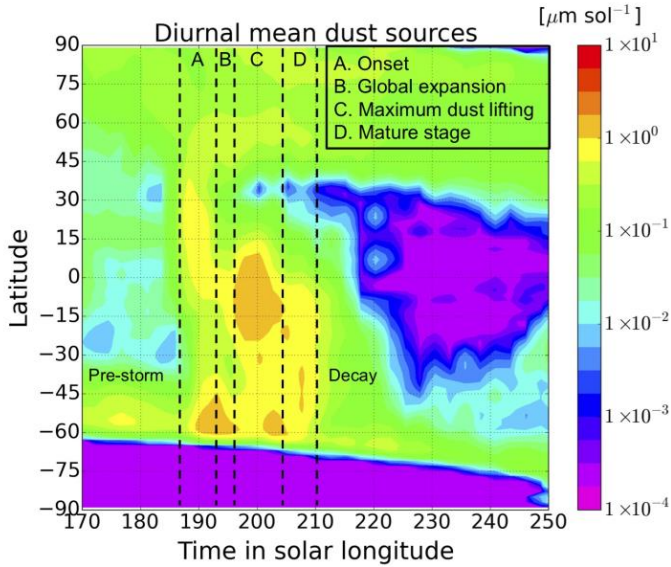

Figure 7: Diurnal and zonal mean sources (i.e., lifting rate) of dust from the reference MY34 GCM simulation. Note the logarithmic scale on the colorbar. The GDS occurs between $L_s=187°-210°$, with mean dust lifting rates greater than 0.3 μm per sol. The dashed black lines indicate the different phases of the GDS: onset, tropical and global expansion, period of intense lifting near Tharsis/Aonia, and mature stage.

## 5 Model Results: Evolution of the MY34 Global Dust Storm

In this section we explore in detail the simulated sources, sinks and pathways of dust during the GDS during each of the four phases identified in Section 4.4. We show the evolution of column dust visible opacity (Figure 8), zonal mean atmospheric dust mass mixing ratio, mass streamfunction, temperatures and zonal winds (Figure 9), and net budget of surface dust (Figure 10) from our reference simulation (see simulation settings in Section 3.3). In addition, Figure 6 details the evolution of dust sources (i.e., lifting rates), atmospheric column dust mass and surface stress including the contribution of each tagged region (as shown by Figure 2). The global evolution of dust sources (Panel B, top) displays three peaks of intense dust lifting occurring around $L_s=191°$, $L_s=200°$ and $L_s=207°$ and two dips of moderate lifting in between, around $L_s=194°$ and $L_s=204°$. These peaks and dips, respectively, correspond to the abrupt and moderate increases of dust opacity and atmospheric column mass shown in the same figure (panel A and panel C, top).

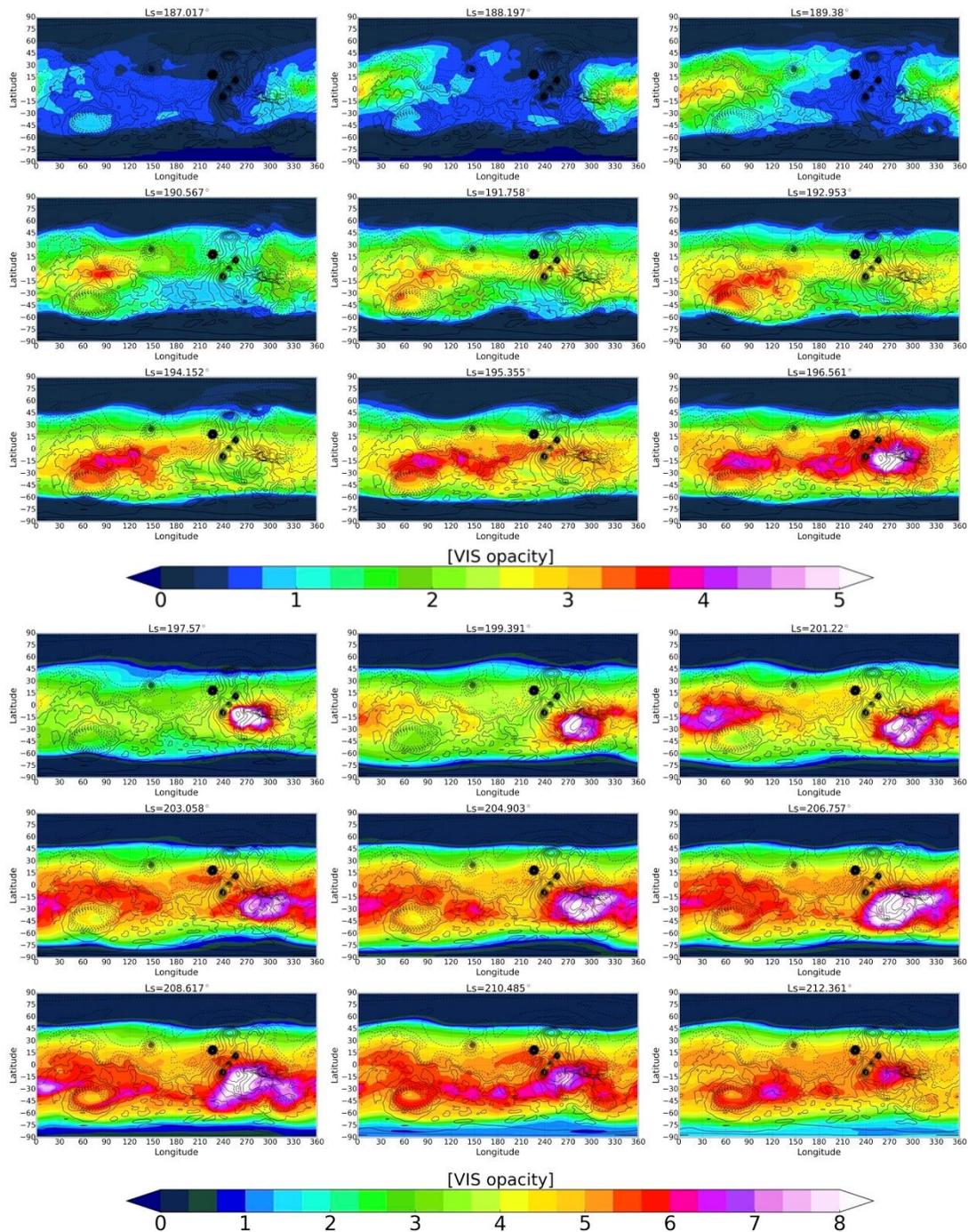

Figure 8: Maps of diurnal mean visible dust opacity (at a reference pressure of 610 Pa) as predicted by our reference simulation every two sols from $L_s=187°$ to $L_s=196°$ (phases A and B, GDS onset and global expansion, top panel), and every three sols from $L_s=197°$ to $L_s=212°$ (phases C and D, maximum dust lifting and mature stage, bottom panel; note the change of scale in the colorbar). The figure shows the development of the regional storm in Acidalia/Chryse and its eastward and southward expansion in Arabia/Noachis, Sabaea/Tyrrhena/Hesperia, before turning into a global storm with $> 5$ opacities at most longitudes in the tropics, and $> 8$ opacities above Tharsis/Aonia.

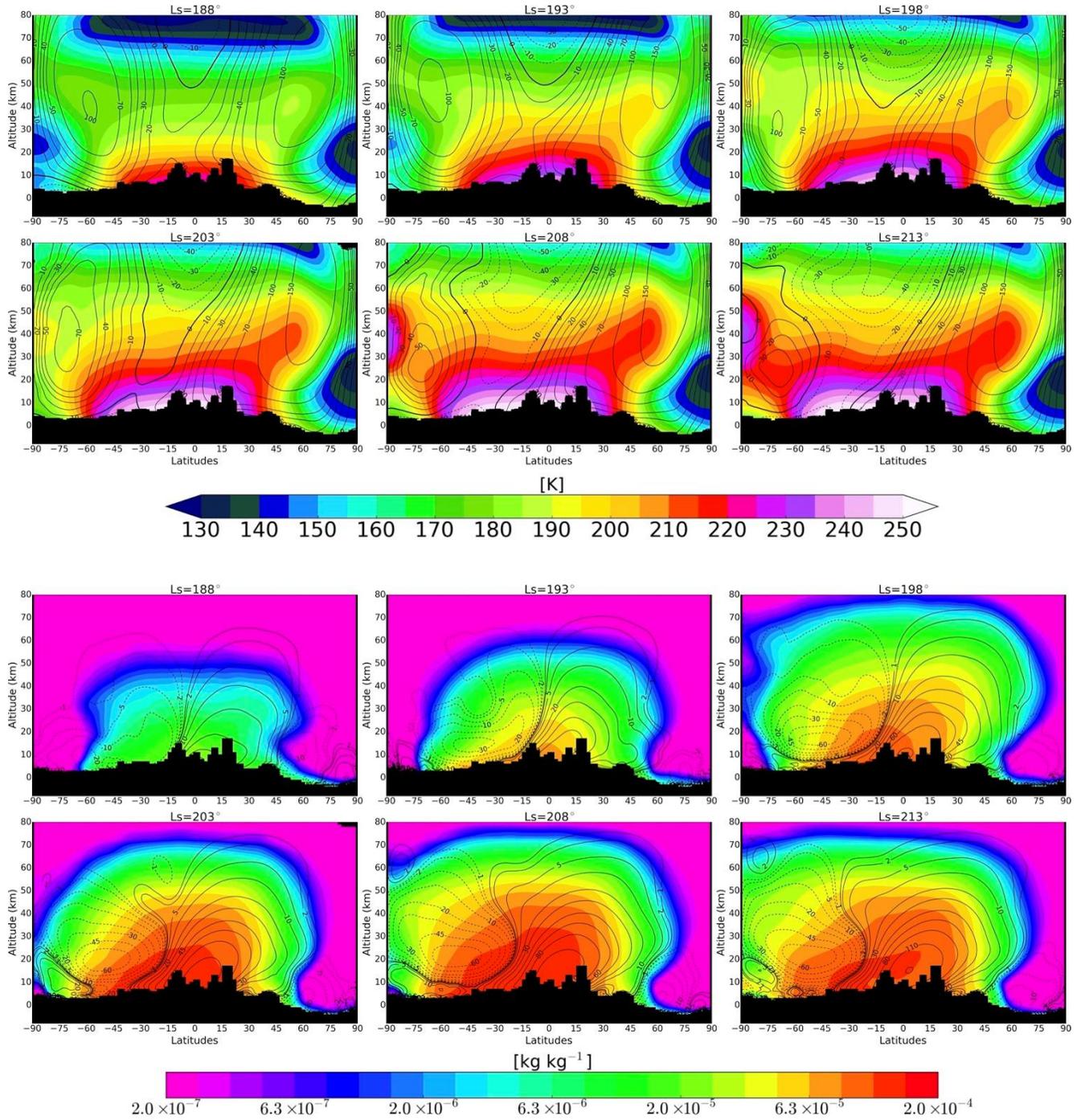

Figure 9: Zonal mean atmospheric temperatures (top panel) and dust mass mixing ratio (bottom panel) averaged over 4 sols a predicted by our reference simulation at roughly 5 sol intervals from $L_s=188°$ to $L_s=212°$. Zonal mean zonal winds are shown with contours in the top panels (a-f) and streamfunction ($10^8$ kg s$^{-1}$) are shown in the bottom panels (g-l), respectively.

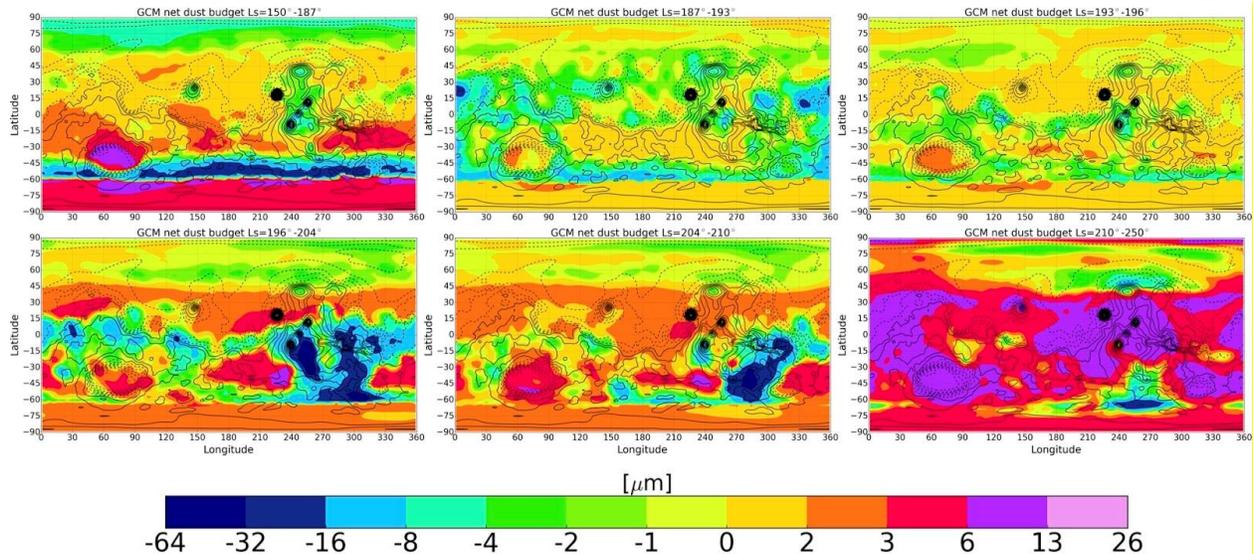

Figure 10: Net surface dust budget as predicted by our reference simulation during the (a) pre-storm period (Ls=150 to Ls=187), (b) onset phase, (c) expansion phase, (d) maximum lifting phase, (e) mature phase, and (f) decay phase ($L_s=210°$ to $L_s=250°$). Hellas and Terra Sirenum are regions where dust tends to accumulate over all these periods. The same applies for Elysium/Acidalia Planitia and the northern plains located between 0°N and 45°N after the onset of the storm.

5.1 Period of Onset of the GDS (Phase A, $L_s=187°$ to $L_s=193°$)

*5.1.1 Dust Sources and Sinks*

During the pre-storm period, $L_s=180°$ to $L_s=187°$, dust lifting in the reference simulation is predicted in the northern plains (Acidalia/Utopia Planitia, Vastitas Borealis) and near the south polar cap between 45°S-60°S. By $L_s\sim187°$, a regional storm forms in Acidalia and develops further south through the Acidalia/Chryse topographic channel. A large and sudden increase of predicted dust lifting occurs between 15°S-30°N in Chryse Planitia and Xanthe/Arabia Terra (Figure 6). This event is responsible for the abrupt increase in dust opacity and temperature observed by MCS (Figure 6). The regional storm grows larger above these regions, with opacities locally reaching 3 (Figure 8). During the same period, lifting along the edge of the retreating south polar cap is still active. The retreating cap edge is simulated to cover latitudes down to at least ~65°S throughout the period of the GDS, thus all dust lifting associated with the cap edge storms occurs at higher latitudes.

By $L_s=190°$, the regional storm moves towards the east to Sabaea/Tyrrhena Terra, around the edges of Hellas and towards the south to Noachis. Intense lifting in these regions produces predicted dust clouds in the general Noachis region to the northwest of Hellas, with peak visible dust opacities of up to 4 (Figures 6 and 7).

Around $L_s=191°$, small amounts of dust are lifted in the Elysium and Tharsis regions, which suggests that the eastward transport of dust from the Arabia/Sabaea region triggered more lifting in those locations by warming the atmosphere, increasing the surface stress and eventually supplying the surface with dust particles that can be lifted again (see below). By $L_s=192°$, the main regional dust storm slightly shifts from Sabaea Terra towards the southeast and grows to cover the entire Noachis/Hellas/Tyrrhena regions, with visible column dust opacities peaking at about 5. The storm also develops east of Hellas, in Promethei Terra, thus merging with cap edge dust lifting. Dust that has been transported southward in Noachis now fills Hellas, with visible column dust opacities of up to 3.5 above the basin (Figure 8).

Figure 10 shows that over the period covering the onset of the GDS ($L_s=187°$ to $L_s=193°$), accumulation of dust onto the surface is predicted in Hellas, Cimmeria/Sirenum, and in Icaria (and above the polar caps, but this remains true all the time as they correspond to permanent sinks of dust), while significant net loss of dust is predicted in the regions where the main regional storm was active (Acidalia/Arabia/Sabaea/Tyrrhena/Noachis) and along the southern polar cap edge.

*5.1.2 Dust Transport: Hadley Cell Circulation and Diurnal Thermal Tides*

Near the south polar cap edge, the return branch of the southern Hadley cell helps in transporting dust lifted near the cap edge towards the tropics (Figure 9), thus allowing the cap edge storms to merge with the main regional storm around Hellas. At the equator, the slow mean convergence of both southern and northern Hadley cells forms a narrow equatorial rising branch (in the zonal mean) which enables vertical expansion of the dust, transporting it up to 50 or 60 km.

Thermal tides add to the Hadley cell circulation to transport dust. The thermal tides are a global-scale atmospheric response to the diurnally varying component of thermal forcing resulting from aerosol heating within the atmosphere and radiative and convective heat exchange with the surface. Their impact on atmosphere temperatures can be seen in Figure 5, which shows a difference of about 30 K between the daytime and nighttime temperatures. The large amount of dust that is injected into the atmosphere by the storm strengthens this contrast, with up to a 80 K temperature difference at high southern latitudes during the mature stage of the storm. The thermal tides have a strong impact on the circulation. Whereas nighttime cooling of the atmosphere leads to atmosphere contraction and downward motion in the tropics, daytime heating leads to atmosphere expansion and upward equatorial motion. Figure 11 (top) highlights the effect of the diurnal thermal tides on the meridional winds as during the onset of the GDS.

Figure 11 (bottom) shows that the meridional winds intensify locally where dust is present. During the onset of the GDS, the tides are enhanced in the "Hellas" hemisphere (0°-180°E) referred, in particular between longitude 0°-60°E where the main regional storm is active and where the atmosphere contains more dust (Figure 8). In the opposite "Tharsis" hemisphere, where dust loading is much less, meridional winds are 10-30 m s$^{-1}$ weaker.

The onset of the GDS and the subsequent increase of atmospheric dust loading heats large parts of the atmosphere because of the absorption of solar radiation by the dust particles. Over the period covering the onset of the GDS ($L_s=187°-193°$), the model predicts an increase of temperatures of up to 10-20 K around 40 km (10 Pa), as shown in Figure 9. The subsequent warming of the surrounding gas causes an expansion of the atmosphere and the strengthening of the thermal tides and of the Hadley cell circulation. This contributes to increasing the surface stress and the dust lifting, leading to a positive amplifying feedback as more dust is injected.

Once dust particles are transported to high altitudes in the equatorial regions of the "Hellas" hemisphere, they are transported eastward by the prograde high-altitude winds (Figure 9), while remaining confined within the narrow corridor formed by the converging southern and northern Hadley cells. Figure 8 shows dust clouds between latitudes 15°S-15°N, in Elysium Planitia and above Tharsis regions during the period Ls=190°-193°, which correspond to high altitude dust transported eastward through this equatorial corridor.

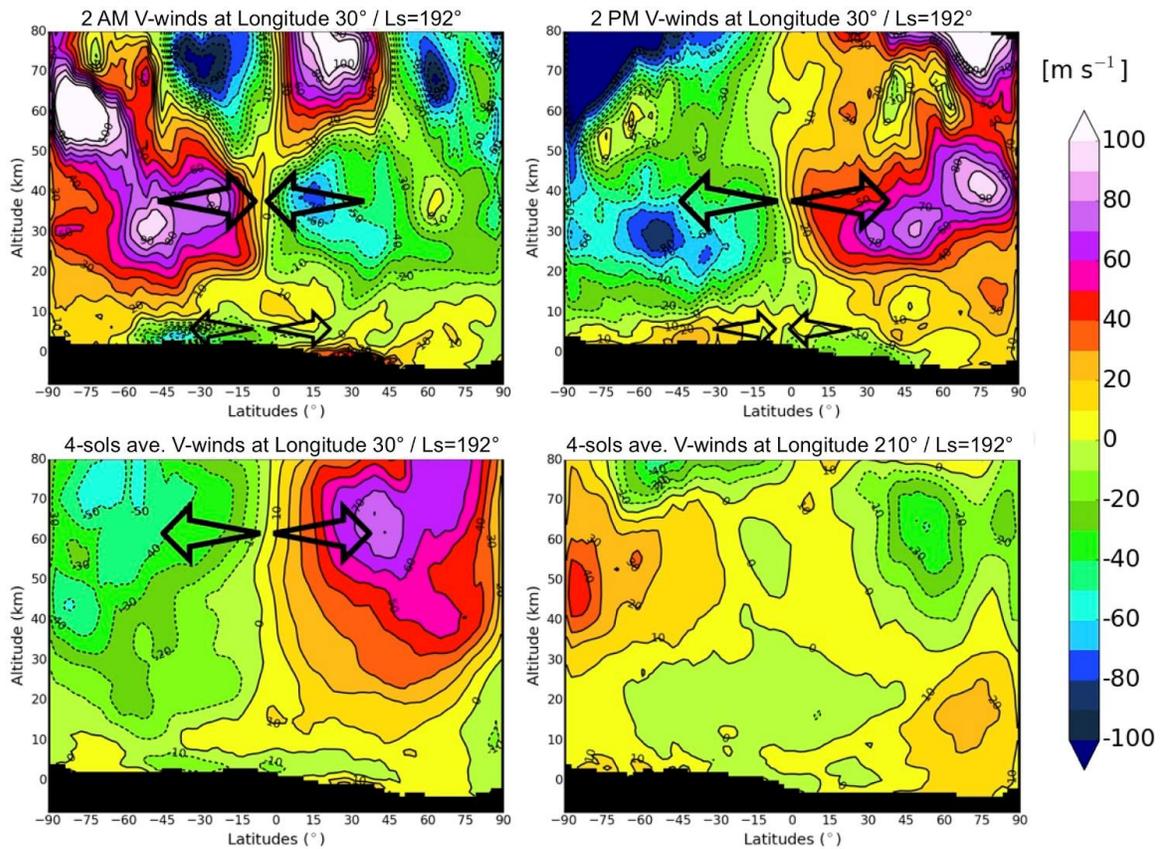

Figure 11: Meridional winds predicted by the reference simulation during the onset of the GDS at $L_s$=192°. Top: 2 am (left) and 2 pm (right) meridional winds averaged over 4 sols at longitude 30° (corresponding to the longitude of the main regional storm and maximum dust lifting at this time), showing the effect of the diurnal tide. Black arrows emphasize the direction of the flow. Bottom: Meridional winds (all times of day) averaged over 4 sols at longitude 30° (left) and longitude 210° (right), showing that the dusty "Hellas" hemisphere produces a stronger meridional circulation than the less dusty "Tharsis" hemisphere does at this time. Note that the diurnally varying meridional winds at low altitudes (below ~6 km) have the opposite phase of the meridional tide at higher altitudes. The tide component is marked by low-level daytime convergence at the equator and divergence at higher altitudes.

5.2 Period of Global Expansion (Phase B, Ls=193°-196°)

*5.2.2 Dust Sources and Sinks*

By $L_s$=193°, the storm is termed global, with dust clouds extending to all longitudes. However, lifting in the "Hellas" hemisphere still dominates, with the main initial regional storm moving eastward and southward, and being mostly active north and west of Hellas (in Tyrrhena Terra and Hesperia/Promethei/Cimmeria), as shown in Figures 6b and 7. This period is marked by a "kink" around $L_s$=194° seen in the observations of the global mean temperatures and opacities (Figure 6a), and reproduced by the model (Figure 6c). This is due to (1) a strong decrease in dust lifting in the Arabia/Xanthe and Noachis regions (Figure 6b), and (2) the fact that the main regional storm lifts less dust in the new regions of active lifting (north and west of Hellas) than it did in the Acidalia/Arabia regions few sols earlier. As a result, the dust lifting remains moderate during this period (in the global average) as shown by reduction in dust lifting rate around $L_s$=194°-195° on Figure 6b, which explains the slower increase in global mean opacity. This "kink" marks the transition between the dust lifting dominating in the "Hellas" hemisphere to dust lifting dominating in the "Tharsis" hemisphere. By $L_s$=195°, dust lifting starts to increase in Tharsis and the southern polar cap edge, leading to a faster increase in opacity (see below). Note that the strong decrease in dust lifting in the Arabia/Xanthe and Noachis regions around $L_s$=190°-195° (Figure 6b) suggests a depletion of surface dust in these reservoirs (subsequent of intense lifting during the onset of the GDS) or lower surface stresses in these regions as the main regional storm moves eastward. This is further discussed in Section 6.

Over the period $L_s$=193°-196°, a net accumulation of dust is obtained in the regions of Cimmeria/Sirenum and Aonia, at the bottom of Hellas, and over a latitudinal band between 0°-30°N (Figure 10). In particular, dust lifted from Arabia/Xanthe and Sabaea/Tyrrhena preferentially accumulates around the Tharsis regions (Figure 12), after being transported eastward through the equatorial corridor. This is interesting because the GCM predicts intense dust lifting in these regions a few sols later, after $L_s$=196° (see below), with quantities of surface dust involved similar to what was accumulated during the previous sols. This suggests that the eastward transport of dust from the "Hellas" to the "Tharsis" hemisphere supplied enough surface dust to the Tharsis regions to trigger significant dust lifting there and maintain the GDS activity.

*5.2.3 Dust Transport*

During this period, dust particles are still transported upward in the tropics and eastward through the narrow equatorial Hadley cells corridor. Figure 9 shows that the Hadley circulation continues to strengthen (thermal tides, not shown, also intensify), in association with stronger equatorial zonal winds, warmer temperatures and a larger vertical extent of dust, reaching up to 70 km altitude. Large dust opacities of about 3-5 are predicted between longitudes 60°E and 300°E, indicating that large amounts of dust have been transferred from the "Hellas" hemisphere, where most of the lifting occurs during this period, to the "Tharsis" hemisphere. At the end of this period ($L_s$=196°), intense dust lifting is triggered over the Tharsis/Solis/Sinai/Syria regions.

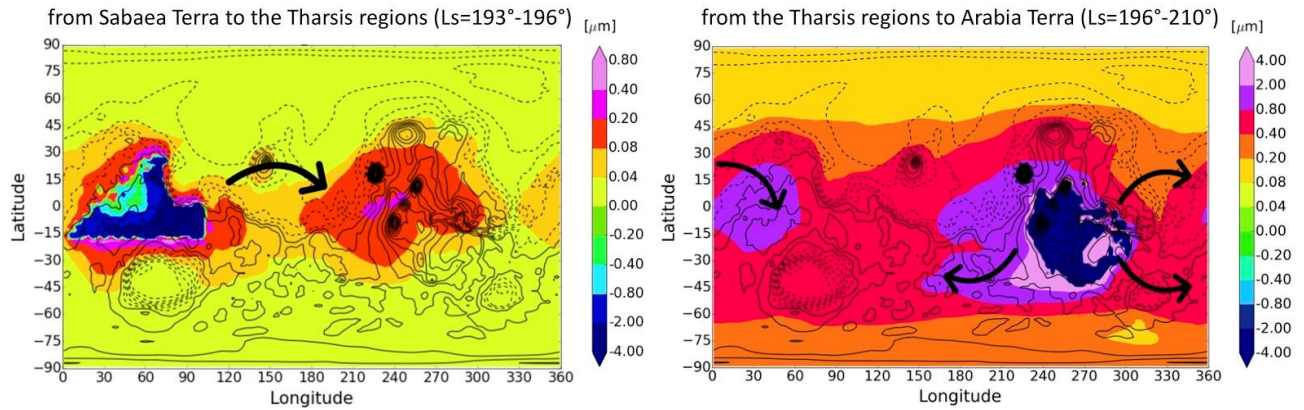

Figure 12: Left: Net budget of surface dust lifted in Sabaea/Tyrrhena (right) over the period ($L_s$ 193° to $L_s$ 196° (a similar pattern applies for dust lifting in Arabia/Xanthe). Dust lifted from these regions preferentially accumulate in the Tharsis region. The black arrows indicate the main pathways of dust. Right: Net budget of surface dust lifted in Tharsis/Syria/Sinai/Solis (left) over the period $L_s$=196° to $L_s$=210°. Large quantities of dust lifted from these regions accumulate in Arabia Terra. Note the non-linear colorbar.

5.3 Period of Maximum Dust Lifting and Large Dust Plumes over Tharsis / Aonia (Phase C, $L_s$=196°-204°)

*5.3.1 Dust Sources and Sinks*

This period corresponds to a maximum in dust lifting in the Tharsis and Aonia regions, associated with a less intense increase in dust lifting in Arabia/Xanthe and Sabaea/Tyrrhena. The dust lifting activity in both eastern and western hemispheres leads to a strong increase in the global mean visible dust opacity, which occurs as rapidly and for as long as during the onset period of the GDS (Figure 6). Between $L_s$=197° and $L_s$=199°, particularly intense and short injections of dust occur in the Tharsis/Syria/Sinai/Solis regions for a few sols, triggering large plumes of dust above these regions (see below). By $L_s$=199°, dust lifting in the Tharsis regions strongly decreases and Aonia Terra dominates most of the dust lifting and the atmospheric column mass of dust by $L_s$=203° (Figure 6).

Around $L_s$=200°-201° (that is, following this intense dust lifting in the Tharsis/Aonia regions), intense dust lifting occurs again in Arabia/Xanthe and Sabaea/Tyrrhena, after the event that occurred around from $L_s$=187° to $L_s$=190° (Figure 6). This second peak of dust lifting in this region may be indicative of significant resupply of surface dust during the $L_s$=190° to $L_s$=200° period. In fact, Figure 12 shows that large amounts of dust lifted from the Tharsis regions during this period of the GDS accumulate in Arabia Terra and its surroundings. This suggests again a connection between both reservoirs of the Arabia/Sabaea and Tharsis/Aonia regions. During this period, the resupply of surface dust from one region to another may play an important role to maintain the GDS activity, especially since large amounts of dust have already been removed from the Arabia/Sabaea regions during the onset of the GDS.

The end of this period ($L_s$=200°-204°) resembles the onset of the storm as dust lifted from Arabia follows similar pathways and is followed by dust lifting in Sabaea, Tyrrhena, Hesperia,

Promethei and Cimmeria (Figure 6). Overall, during this period of the GDS, net loss of dust is obtained in Tharsis/Aonia, Arabia/Xanthe and Sabaea/Tyrrhena/Hesperia regions (Figure 10). Note that cap edge lifting is still significant. A net accumulation of dust is predicted in Hellas, Sirenum/Icaria, Noachis, the south pole and a latitudinal band between 0°-45°N including the low topographic regions of Elysium/Acidalia/Amazonia/Chryse and north of Arabia Terra.

*5.3.2 Dust Transport: Large Plumes of Dust over the Tharsis and Aonia Regions and Changes in the Zonal Transport of Dust*

The intense surface dust lifting predicted during the period $L_s=197°$ to $L_s=199°$ leads to large amounts of dust injected into the PBL, which produces high local dust opacities. Figure 13 shows the diurnal evolution of atmospheric dust mass mixing ratio over four sols during this period, but only the dust tagged as being lifted from Solis/Sinai Planum at $L_s=198°$, where the most intense dust injection is predicted. The solar heating of this thick layer of dust near the surface drives a strong atmospheric heating anomaly and subsequent rapid and dramatic vertical motions, transporting dust up to 70 km during daytime (12 pm - 6 pm). This event is associated with local maximum column visible dust opacities of up to 10 (Figure 8) and warmer atmospheric temperatures above Tharsis/Solis/Sinai/Syria (Figure 5). The mechanism is comparable to the case of the "solar escalator" (*Daerden et al*., 2015) and the resulting plumes strongly resembles the "rocket dust storms" (*Spiga et al*., 2013), but at a much larger scale as the plumes of dust cover up to 60° of longitude (rocket dust storms specifically refer to mesoscale events). During nighttime, the dust plume dissipates, sediments, and is transported downward as it cools off. Subsequent detached layers of dust are obtained between 20 and 60 km altitude and are transported mostly eastward by the high-altitude winds, and westward by the near-surface winds. Figure 13 shows how atmospheric dust is impacted by the diurnal thermal tides, as it is transported upward and downward over tens of kilometers during daytime and nighttime, respectively. After four sols, dust lifted from the Tharsis regions reaches the "Hellas" hemisphere and mostly accumulates in Arabia/Sabaea (see above).

By $L_s=200°$, the Hadley circulation transitions to one cell with a meridional flow dominated by a circulation from the summer to the winter hemisphere, as shown by Figure 9. The zonal circulation in the equatorial regions becomes weaker, while the prograde jet strengthens in the winter (northern) hemisphere and weakens in the summer (southern) hemisphere. As a result, the large plumes of dust formed in the equatorial region near Tharsis are not transported zonally as fast as during the previous periods of the GDS. Near-surface and high altitude equatorial dust is transported westward by slow retrograde winds (Figure 9), while dust reaching higher latitudes are transported eastward by the northern and southern prograde jets.

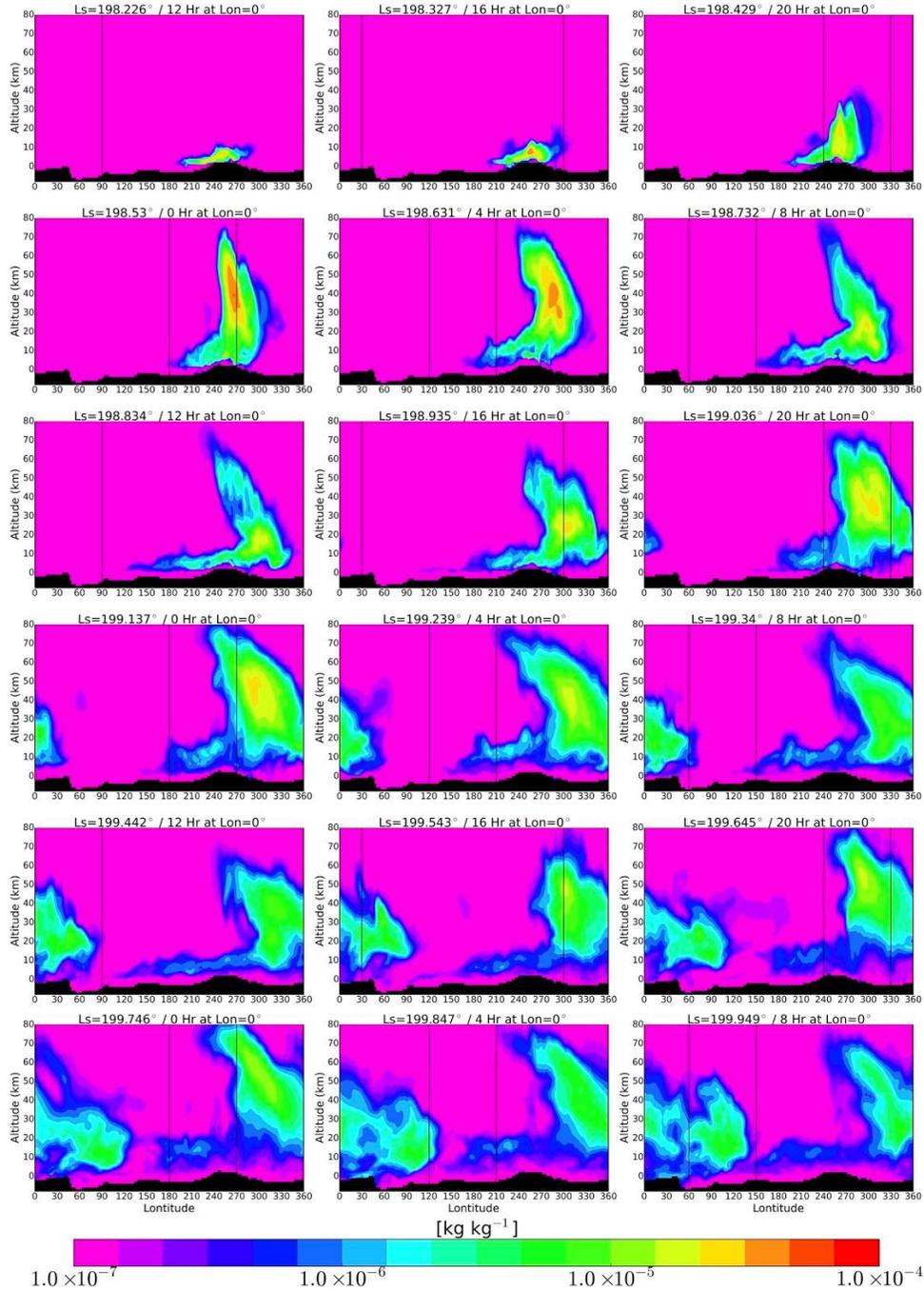

Figure 13: Longitudinal cross-section of the atmospheric dust mass mixing ratio averaged over latitudes 45°S to 40°N, during four sols between $L_s=198°$ and $L_s=200°$ at four different local times longitude 0 (snapshots at 12 am, 6 am, 12 pm, and 6 pm). Only the tagged contribution of dust lifted from Solis/Sinai Planum at $L_s=198°$ is shown (see location on Figure 2), in order to highlight and track the plume of dust subsequent to the intense lifting occurring at this time and location.

5.4 Mature Stage of the Storm (Phase D, $L_s=204°$-$210°$) and Decay

During the mature stage of the GDS ($L_s$=204°-210°), dust lifting is intense in Aonia. Large plumes of dust, comparable to those shown in Figure 13, are still predicted in this region, with local opacities reaching ~10 (Figure 8). Dust lifting is limited in other regions (Figure 6). On a global scale, the sources of dust still dominate the sinks (dust particle gravitational sedimentation), which is why the global opacity continues to increase during this period. The peak of the global mean dust opacity is reached at $L_s$=209°, due to dust lifting weakening in most regions, including Aonia Terra, and to a reduced area of dust lifting in general, thus triggering the decay phase of the GDS.

A net loss of dust in Tharsis/Aonia (and to a lesser extent in some regions around Hellas and in the northern plains) is predicted during this period (Figure 10). As during the previous period, net accumulation of dust is obtained in Hellas, Sirenum/Icaria, Noachis, Hesperia/Promethei, the south pole and a latitudinal band between 0° N and 45°N including the low topographic regions of Elysium/Acidalia/Amazonia/Chryse and north of Arabia Terra. During the period of maximum global mean opacity at $L_s$=209°, the total amount of dust mass removed from the surface in the simulation (compared to its level at $L_s$=180°) is about 0.012 kg m$^{-2}$ (~4.8 microns; Figure 6). Atmospheric temperatures and dust loading are maximum during this period (Figures 5 and 9).

During the decay phase of the storm, most of the airborne dust is sourced from previously active lifting centers such as the Tharsis and Aonia regions. Large quantities of dust fall back to the surface, mostly in the southern hemisphere between 60°S and 30°S, and in particular in Tharsis, Elysium, Hellas and Arabia Terra (Figure 10). The Hadley cell circulation decreases in intensity with the decrease of atmospheric dust loading, while the westward flow in the southern hemisphere strengthens.

## 6 Discussion

In this section, we summarize what we have learned about the MY34 GDS and discuss our results in light of comparisons with the MY25 GDS, regional storms, and non-dusty conditions.

### 6.1 Hadley Cell Circulation and Thermal Tides

*6.1.1 The Importance of the Zonal Circulation and Thermal Tides*

Our modeling of the MY34 GDS highlights the rapid zonal expansion of the storm during its onset ($L_s$=187°-193°). The strong equinoctial season eastward jet in the tropical regions efficiently transports dust from the "Hellas" to the "Tharsis" hemisphere. In particular, we show a significant net transport of dust from the active regions of dust lifting at this time, Chryse/Arabia, to the Tharsis regions (see Figure 12). This eastward expansion of the storm remains at first relatively confined to the equatorial regions, due to the converging lower branches of the northern and southern Hadley cells which sweep dust into the upward circulation at equatorial latitudes. As the most active regions of dust lifting are located near the equator during this time, large amounts of dust are transported eastward, creating a "corridor" of high-altitude dust above the equator (see Figure 8). In addition, descending motion in the upper branches of the Hadley cells tend to confine dust lifted to higher latitudes near the surface and transport it toward the equatorial region where it is then easily transported vertically to higher altitudes.

Figure 14 shows the zonal wave 1 component of the $T_{15\_diff}$ fields for the MY34 GDS, evaluated in the southern subtropics (20°S). This field represents the zonal asymmetry in the depth-weighted diurnal temperature range that is expected to be strongly driven by aerosol forcing. This measure succinctly summarizes a key aspect of the development of the storm, which is the eastward progression of dust lifting. The MY34 storm appears to have developed in two discrete stages, which are well reproduced by our simulations: (1) The initial storm development at $L_s$=187° as a flushing storm migrated across the equator at 330°E (includes our phase A and B: the onset and global expansion of the storm), and (2) A second phase of the storm expansion with significant lifting begin triggered in the Tharsis region (includes our phases 3 and 4).

In this paper, we also highlighted the strong effect of diurnal thermal tides on dust transport. Diurnal thermal tide forcing is usually strongest (with highest temperature amplitude) during equinox, and it strengthens when large amounts of atmospheric dust are present in the tropics, as it is the case for the GDS [*Wilson*, 2012a; 2012b; *Barnes* et al. 2017]. Here our simulations show that large plumes of dust are transported downward during nighttime and upward during daytime over tens of km by the thermal tides. The daytime upward transport of dust dominates and largely contributes to the injection of dust to high altitudes and to its zonal transport, as plumes of dust are able to remain for a longer time in the atmosphere. In addition, the daytime expansion of the atmosphere driven by the thermal tides allows for significant meridional transport of dust (although at a much slower rate than the zonal transport of dust).

Figure 15 shows pronounced differences in the height and meridional extent of the simulated aerosol field between daytime and nighttime that is due to the advecting influence of the sun-synchronous diurnal-period thermal tide. The predicted maximum southward dust extent occurs around 1800 LT and is in phase with the simulated phase of the diurnal period temperature variation in the southern hemisphere. Maximum poleward tide amplitude for V (up to 200 m/s) occurs at ~10 Pa around midday.  Similar variations in dust cloud height and extent are seen in MCS limb retrievals during the MY34 storm [Armin Kleinboehl, personal communication]. The upward advection of dust in the tropics is also evident.

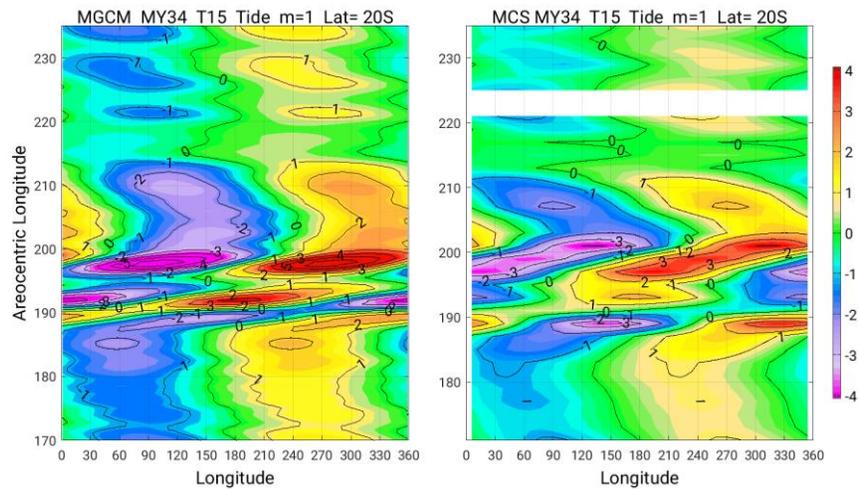

Figure 14: (a) The zonal wave 1 component of the MY34 tide field $(T_{pm}-T_{am})/2$ at 20°S based on MCS T15 observations, binned at 2° of $L_s$. The field is evaluated with 5 sol averages. (b) The zonal wave 1 component of the MY34 tide field $(T_{pm}-T_{am})/2$ at 20°S based on $T_{15}$ as simulated in our reference simulation.

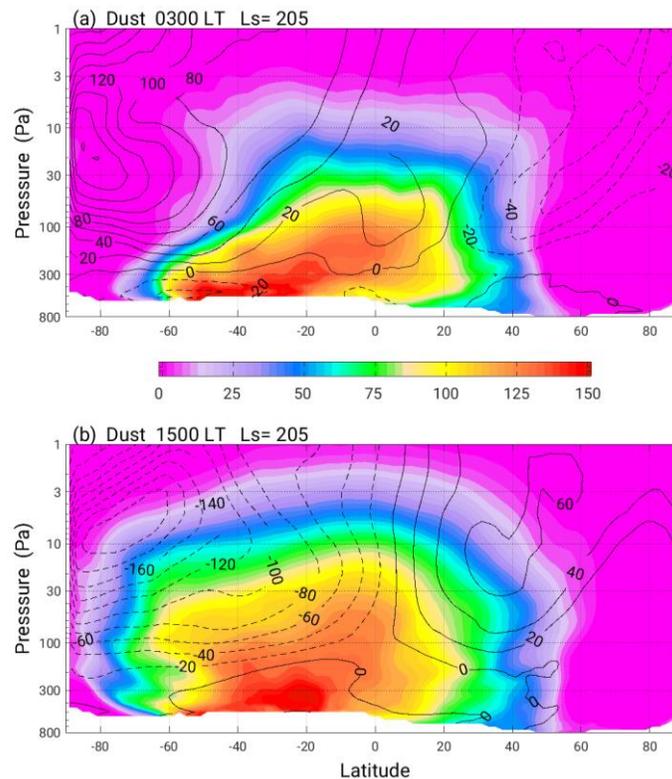

Figure 15: (top) Simulated zonally averaged 3 am aerosol mass mixing ratio for $L_s$=205°. Units are ppm. The zonally averaged 9 pm meridional wind field is contoured at 20 m s$^{-1}$ intervals. (bottom) Simulated zonally averaged 3 pm aerosol mass mixing ratio for $L_s$=205°. Units are ppm. The zonally averaged 9 am meridional wind field is contoured at 20 m s$^{-1}$ intervals.

*6.1.2 Sensitivity of the Hadley Cell Circulation to Dust Loading and Dust Properties*

The increase in atmospheric dust loading and subsequent heating of the atmosphere leads to a higher thermal contrast between the equatorial and polar regions resulting in a more intense mean meridional circulation, a more extended Hadley cell circulation, and a positive amplifying feedback as more dust is injected into the atmosphere from increasing surface stress lifting.

This is in agreement with previous modeling of Mars's atmosphere showing an increase of the intensity of the Hadley cell circulation with higher dust loading [*Basu et al*., 2006; *Wilson*, 2012a]. This is highlighted in Figure 16, which shows a direct comparison of the general circulation (mean meridional and zonal circulation, atmospheric temperature and dust mixing ratio) obtained with the MGCM during the MY34 GDS (reference simulation) and during MY33 at a similar season. The MY33 simulation was performed with same settings as the reference simulation.

The intensity of the Hadley circulation in our reference simulation is relatively strong compared to other modeling results of the Martian dynamics in dusty conditions for this season. In particular, we find a meridional mass flux of up to $60 \times 10^8$ kg/s at ~20 km in altitude in both the northern and southern Hadley cells. The intensity of this circulation is sensitive to the dust particle size distribution, whereas it remains relatively insensitive to the other simulation parameters listed in Section 4.3 (sensitivity study). The maximum meridional mass flux decreases by a factor 2 when decreasing the reference radius from 3 µm to 1.5 µm. This is because increasing particle size significantly decreases the ratio of dust opacity in the visible compared to the IR [*Murphy et al*., 1993]. This is a key parameter in the model since it controls the local radiative balance of the atmosphere and the amplifying positive feedback that dust particles have on the general circulation. *Wilson and Hamilton* [1996] reported an increase in Hadley circulation intensity and thermal tide amplitude with an increase in IR emissivity. *Haberle et al*. [1997] used an idealized Mars atmosphere model with Newtonian cooling for heating to show that Hadley circulation intensity was inversely proportional to the thermal damping timescale. Of course, the Hadley circulation continues to intensify as the season advances towards the solstice [*Basu et al.,* 2006; *Haberle et al*., 2018]. Nonetheless, the dust storm continues to decay despite this intensification, pointing out that negative feedbacks must come to the fore to explain the finite lifetime of the GDS.

*6.1.3 Impact of the Circulation on Water Vapor and Water Ice Clouds*

Although this paper focuses primarily on the evolution of the dust storm and the pathways of dust, we have investigated the behavior of water vapor and ice particles in our reference simulation. Recent observations of the water vapor abundance in the Martian atmosphere have been reported during dust storm conditions and revealed an increase in high-altitude atmospheric water vapor at high northern latitudes [*Fedorova et al*., 2018, *Heavens et al*.,2018, *Vandaele et al*., 2019]. Figure 16 shows the zonal mean distribution of water vapor and water ice in the reference simulation during the GDS expansion, and compares the result with that obtained with the same GCM simulation but performed during MY33 (less dusty conditions). During the GDS period, because of the higher temperatures, fewer water ice clouds are predicted, and the remaining ice particles form at higher altitudes. As a result, the atmosphere is enriched in water vapor, which is efficiently transported to high altitudes, along with dust particles, as a result of atmosphere heating and expansion. By contrast, under the non-storm conditions of MY33, water ice clouds particles form at lower altitudes and confine water vapor down lower as ice particles sediment. This is consistent with the recent water vapor profiles retrieved during the MY34 GDS and with previous numerical modeling of regional dust storms, which demonstrate that water vapor would be transported to higher altitude along with dust by the large-scale upward motions driven by daytime solar heating of the dust layers [*Daerden et al.,* 2015, *Spiga et al.*, 2013].

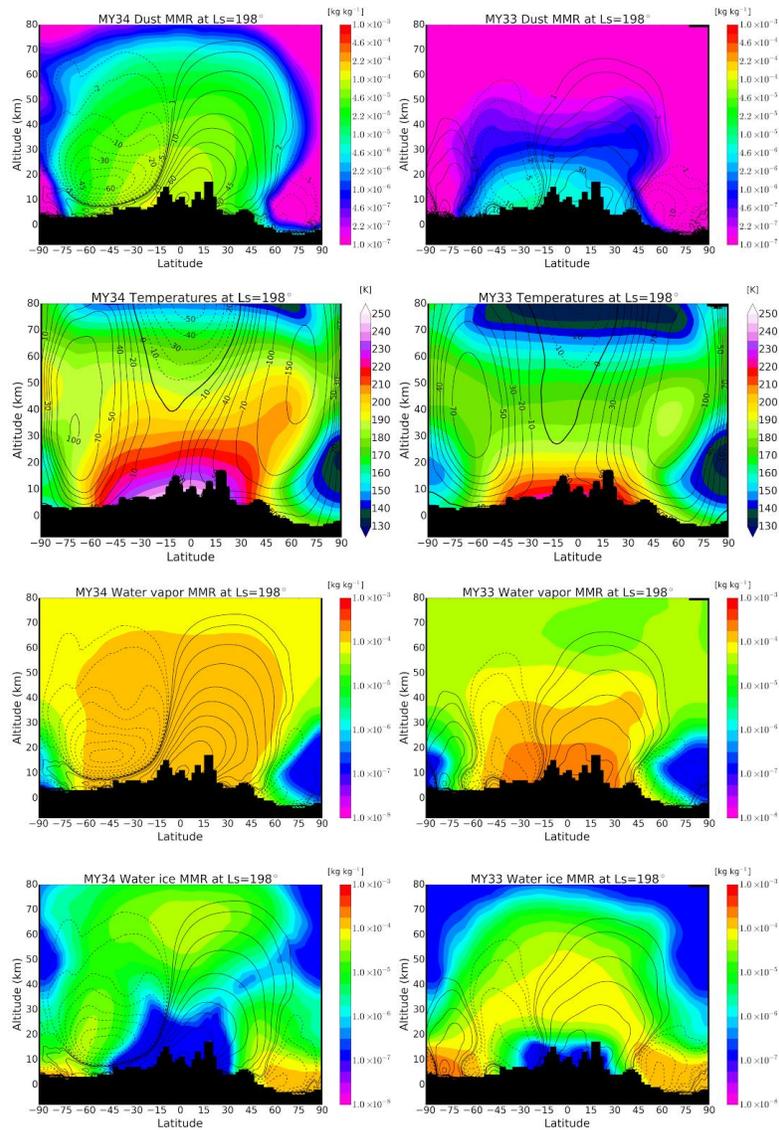

Figure 16: Comparison between MY34 (left) and MY33 (right) GCM simulations at $L_s$=198°: 4-sol averaged zonal mean of (A) atmospheric dust mass mixing ratio with streamfunction contours, (B) atmospheric temperatures with zonal wind contours, (C) atmospheric water vapor mixing ratio and (D) water ice mixing ratio.

6.2 Dust Sources and Sinks: Comparisons with the MY25 GDS

*6.2.1 Overview: Similarities and Differences between MY25 and MY34 GDSs*

The MY25 and MY34 GDSs have a nearly identical timing of onset ($L_s$=185° for MY25 vs $L_s$=187° for MY34) and dust storm peak (between $L_s$=205° $L_s$=210°) as well as similar declines in opacity and atmospheric temperatures (Figure 1), though it appears that localized dust lifting in Solis/Syria/Sinai Planum persisted longer in MY25 (out to $L_s$=225°, *Cantor et al.*, [2007]) than in MY34. The occurrence of both storms is in advance of the regular "A season" storms [*Kass et al.,* 2016], which range from $L_s$=208° to 234°. However, the MY25 GDS storm was initiated in the southern hemisphere, north of Hellas (Hesperia Planum). During MY25, dust was transported eastward along the equator and evidently subsequent lifting was triggered in Daedalia/Sinai/Solis Planum, which led to very widespread lifting over a broad range of latitudes (extending to the cap boundary). The MY34 event was followed by a large "C season" event (evidently the strongest to date), while the MY25 "C season" storm was relatively minor.

*6.2.2. Dust sources and Sinks during the MY34 and MY25 GDS*

Figure 17 shows the net budget of surface dust obtained in the simulation during the MY34 GDS period ($L_s$=180°-250°) compared to a simulation of the MY25 GDS (performed with similar settings to our reference simulation). Over the MY34 GDS, a net accumulation of surface dust is predicted in Hellas, Cimmeria/Sirenum Terra (30-40 µm) as well as in northern Arabia Terra, Elysium/Amazonis, Hesperia, Noachis and the south polar cap (10-20 µm). A net loss of dust is obtained in Sabaea/Tyrrhena (20-30 µm), in Aonia/Tharsis/Syria and along the Thaumasia Plateau (60-80 µm), in Xanthe/Lunea (30-40 µm) and along the south polar cap at latitude 60°S (20-40 µm). The results obtained for the MY25 GDS show similar patterns, with a significant loss of surface dust in the Tharsis/Syria and Sabaea/Tyrrhena regions, and along the polar cap edges. This reinforces the global view that both MY25 and MY34 GDS have similar evolution and pathways.

We note that there is significant dependence in the details of cap edge lifting in both MY34 and MY25 simulations. This may be related to differences in the dust scenarios, which are derived from TES and MCS observations for MY25 and MY34, respectively. Near the cap edge, whereas MCS limb observations would tend to underestimate the dust opacity (because dust is confined near the surface), TES nadir observations would tend to overestimate the dust opacity because of the complicating influence of patchy surface frosts, and the weak contrast between surface and air temperatures [*Montabone et al.,* 2015].

It would be interesting to compare the maps of net budget of surface dust obtained with the GCM with maps of observed albedo changes covering the MY34 GDS period. We anticipate that a darker surface albedo will be seen in the Tharsis/Syria/Aonia regions as large amounts of dust has been removed from the surface in these regions, according to the model. Conversely, we predict brighter regions around Hellas, in Sirenum, Hesperia and Elysium/Amazonis. While maps of albedo changes are not yet available for MY34, a first comparison can be made with the maps showing the year-to-year MY25-MY26 differences of surface albedo as observed by TES, published in *Swzast et al.* [2006] (see Fig. 3 and 4) and *Smith et al.* [2004] (see Fig. 14). These maps could largely indicate where dust has been transported during the MY25 GDS, and provide hints of dust redistribution for the MY34 GDS, assuming that both MY25 and MY34 storms have relatively similar evolutions (which is what our simulations suggest).

Our maps of net surface dust budget are partially in agreement with the observed albedo changes for MY25. According to these maps of MY25-MY26 albedo changes for MY25, the Martian surface significantly darkened in the regions of Tharsis/Thaumasia Plateau, including Syria/Sinai. This is in good agreement with our simulations, although the observed surface darkening is more extended that the simulated net loss of dust. In addition, dust accumulation is predicted by the GCM in Cimmeria/Sirenum, Hellas, Tyrrhena/Hesperia, northern Arabia Terra, and in the northern low-topographic plains below 45°N, which is in agreement with the observed albedo changes. However, the GCM predicts strong cap edge lifting for MY25, which is not suggested by the albedo maps. In the regions of Xanthe, Solis and south Arabia Terra, a net dust accumulation is suggested from the observed TES albedo changes during MY25 whereas a net loss of dust is predicted from both simulations of the MY34 and MY25 GDS.

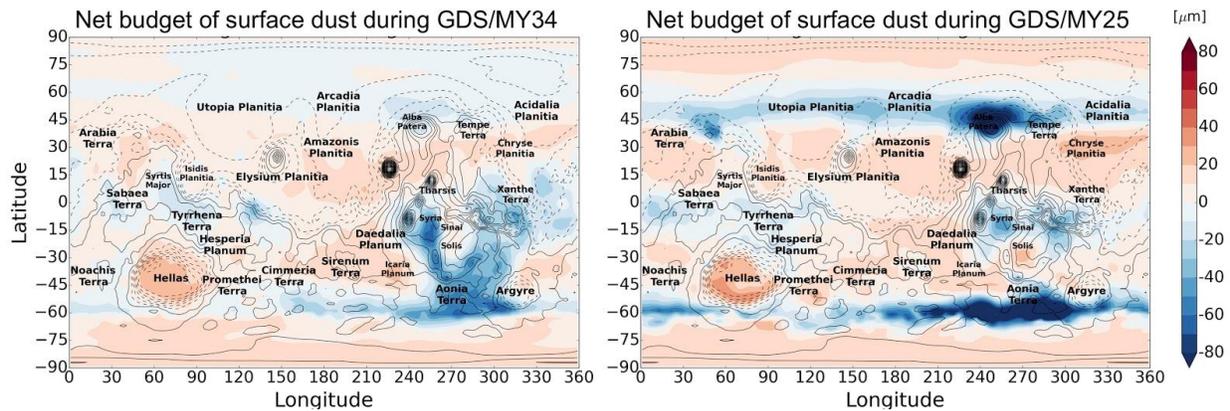

Figure 17: Map of the net budget of surface dust predicted by the GCM for the MY34 (left) and MY25 (right) GDSs from $L_s$=180° to $L_s$=250°.

6.3 Evolution of the MY34 GDS

*6.3.1 Equilibrium Between Surface Stress and Available Dust*

On Mars, dust lifting and the subsequent formation of dust storms could be controlled by an equilibrium between surface stress and available dust. In order to go out of equilibrium and trigger the formation (or decay) of a regional or global dust storm, the surface stress and/or the availability of dust must increase (or decrease, respectively). Here our simulations suggest that both mechanisms can play a role in triggering the formation and the decay of the MY34 GDS.

First, during the GDS onset period, the rapid zonal transport of dust, lifted from the "Hellas" hemisphere, could quickly supply the surface of the "Tharsis" hemisphere with available dust. This scenario is favored to explain the increase of dust lifting in the Tharsis regions around $L_s$=193°, because it is subsequent to a significant resupply of dust in this region (Figure 12). Alternatively (or in addition), the rapid transport of dust from both hemispheres could rapidly increase surface stress and trigger additional lifting centers, in Tharsis for instance. This is difficult to assess with our simulations, because the dust lifting (and subsequent increase of surface stress) is controlled by the prescribed dust opacity maps.

Second, during the period of maximum dust lifting (around $L_s$=196°), increased dust lifting is predicted in the Arabia/Sabaea regions. This is surprising because large amounts of dust have already been lifted from these regions during the onset of the storm, suggesting a depletion in surface dust (Figure 6d). In addition, whereas maximum surface stress increases in most of the tagged regions during the GDS period, the maximum stress in Arabia/Sabaea precisely decreases after $L_s$=195°. This suggests that the increase in dust lifting in the Arabia/Sabaea region after $L_s$=196° is due to a resupply of surface dust rather to an increase of surface stress. This is consistent with our simulations showing a net transfer of surface dust from the Tharsis regions to the Arabia/Sabaea at this time (Figure 12).

Finally, the GDS decays around $L_s$=210°, while the simulated maximum diurnal stresses are still globally increasing. This suggests again that the availability of surface dust plays a role in ending the storm. In fact, at this time, the area of dust lifting seems to be reduced to the Aonia/Tharsis regions. The zonal circulation is very weak as eastward flows change over to westward flows in the southern hemisphere, and thus zonal transport of dust is not efficient. It is possible that the Aonia/Tharsis regions become depleted of surface dust around $L_s$=210° and that reservoirs at other longitudes have not been replenished in surface dust enough (because of the weak zonal circulation) to maintain significant lifting. As a result, dust sedimentation dominates dust lifting and the GDS decays.

### 6.3.2 Comparisons with Regional Storms and other GDSs

Pre-solstice and post-solstice regional storms (also referred to as A and C storms respectively, *Kass et al.* [2016]) are characterized by rapid meridional migrations from high latitudes to equatorial regions through the low topographic channels of Acidalia/Chryse or Utopia/Isidis. They never occur near equinox. "A" storms have never been observed earlier than $L_s$=208°. At this time tropical zonal winds are moderate, and as dust is transported through the tropics, it is not efficiently transported zonally and remain relatively confined in longitude. Consequently, the processes governing these regional storms and the MY34 GDS appear to be distinct.

The MY34 GDS appears to have developed from a flushing storm sequence originating in the northern hemisphere. These frontal storms may have brought the right amount of available dust in the tropics, and in particular in Acidalia/Chryse/Arabia at locations where high surface stress occurs. The subsequent regional storm could have been sufficiently large to cross some threshold of dust availability, transportability, and surface stress, so that it activated lifting centers in the "Hellas" hemisphere (Sabaea, Hesperia, Noachis). In addition, the timing of this event, occurring near equinox, is favorable for a zonal (eastward) transport of dust by the tropical westerlies that are present in a relatively narrow seasonal interval around equinox.

GDSs seem to occur either near southern spring equinox (MY25, MY34) or near southern summer solstice (MY28), or even at both within the same year (MY12). The early season MY12 global storm observed by Viking began at $L_s$~205°, but its development was not well observed. Their development may be related to rapid zonal expansion in the tropics (eastward during equinox, westward during solstice), which would perhaps allow for greater amounts of dust to be quickly lifted than during the A and C storms. In fact, if the right combination of dust availability and surface stress for intense lifting occurs late after equinox or late after solstice (the A and C seasons), the zonal circulation may be too weak leading to the more north-south oriented flushing A and C storms.

Whereas equinoctial GDSs can originate either from dust lifting in the northern (MY34) or in the southern (MY25) hemisphere (but with subsequent significant dust lifting near the equatorial regions), solsticial GDSs may only originate from dust lifting in the southern summer hemisphere because the descending branch of the Hadley cell in the winter northern hemisphere would confine dust near the surface so that its zonal transport is limited. This was the case for the MY28 GDS, as there was a flushing event in Chryse immediately prior to the MY28 GDS that originated in Noachis [*Wang and Richardson*, 2015].

# 7 Summary and Conclusions

In this paper, we simulated the MY34 GDS with the NASA Ames Mars GCM which includes an assimilated dust lifting forced by MCS-derived column dust opacity maps. Our results are in a generally good agreement with the available observations of the GDS, including the atmospheric $T_{15}$ temperatures and the column dust opacities. Although some discrepancies between our simulated dust opacities and those observed are obtained locally, our GCM results remain robust as demonstrated by a sensitivity study showing no significant changes in dust transport, sources and sinks.

During the storm onset and development, we find that dust is efficiently and rapidly transported upward and eastward by the Hadley cell and diurnal thermal tide circulation, in particular in the equatorial regions. Both the Hadley cell and the thermal tides increase in intensity as more dust is injected into the atmosphere due to a positive-radiative feedback. In particular, the model predicts large plumes of dust during the mature stage of the storm, resembling planet-scale rocket dust storms, and injecting dust at high altitude up to 80 km. As a result of atmospheric warming in response to dust heating, we also find that the water ice cloud condensation level migrates to higher altitudes, leading to enrichment of the upper atmosphere in water vapor.

We highlight significant rapid back and forth transfers of surface dust occurring during the development of the storm between reservoirs located in the "Hellas" (e.g. Arabia/Sabaea) and "Tharsis" hemisphere, which result from the rapid zonal circulation characteristic of this storm. These exchanges of surface dust may play an important role in the storm development as they allow for fast replenishment of the surface with available dust.

In our simulations of the storm, the intensity of the Hadley cell is significantly stronger than that in non-dusty conditions. We find the intensification to be strongly sensitive to the radiative properties of dust particles that is tied to the lifted dust particle effective radius.

In the future, we plan to use our sensitivity study to investigate how to realistically refine the observation-derived column dust opacities maps. In addition, we plan to explore the GDS with multi-modal dust distributions, or with modified dust scenario in order to better reproduce and understand the onset and decay of the storm.

**Acknowledgments**

T.B was supported for this research by an appointment to the National Aeronautics and Space Administration (NASA) Post-Doctoral Program at the Ames Research Center administered by Universities Space Research Association (USRA) through a contract with NASA. Electronic output from all model simulations will be made available to the public through the NASA

Advanced Supercomputing (NAS) data portal (https://data.nas.nasa.gov/), which is housed at NASA Ames Research Center.

## References

Arya, S.P. (1988). Introduction to Micrometeorology, Academic, San Diego, CA.

Basu, S., Richardson, M. I., Wilson, R. J. (2004), Simulation of the Martian dust cycle with the GFDL Mars GCM, Journal of Geophysical Research, 109, E11.

Basu, S., Wilson, R. J., Richardson, M. I., Ingersoll, A. (2006), Simulation of spontaneous and variable global dust storms with the GFDL Mars GCM, Journal of Geophysical Research, 111, E9.

Bertrand, T., Kahre, M. A., Wilson, R. J. (2018), Tagging Dust and Water in the NASA Ames Mars GCM : a New Global Vision of the Current and Past Martian Climates, American Geophysical Union, Fall Meeting 2018.

Cantor, B. A. (2007), MOC observations of the 2001 Mars planet-encircling dust storm, Icarus, 186, 1, p. 60-96.

Cantor 2019, this special issue

Clancy, R. T., Wolff, M. J. and Chrstensen, P. ,R. (2003), Mars aerosol studies with the MGS TES emission phase function observations: Optical depths, particle sizes, and ice cloud types versus latitude and solar longitude, J. Geophys. Res., 108, Issue E9.

Daerden, F., Whiteway, J. A., Neary, L., Komguem, L., Lemmon, M. T., Heavens, N. G., Cantor, B. A., Hébrard, E., Smith, M. D. (2015), A solar escalator on Mars: Self-lifting of dust layers by radiative heating, Geophysical Research Letters, Volume 42, Issue 18, pp. 7319-7326.

Fedorova, A., Bertaux, J.-L., Betsis, D., Montmessin, F., Korablev, O., Maltagliati, L., Clarke, J. (2018), Water vapor in the middle atmosphere of Mars during the 2007 global dust storm, Icarus, Volume 300, p. 440-457.

Greybush, S. J., Wilson, R. J., Hoffman, R. N., Hoffman, M. J., Miyoshi, T., Ide, K., McConnochie, T., Kalnay, E. (2012), Ensemble Kalman filter data assimilation of Thermal Emission Spectrometer temperature retrievals into a Mars GCM, Journal of Geophysical Research, Volume 117, Issue E11.

Guzewich, S.D., R.J. Wilson, T.H. McConnochie, A.D. Toigo, D.J. Banfield, and M.D. Smith, (2014). Thermal tides during the 2001 martian global-scale dust storm. J. Geophys. Res..

Guzewich, S. D., Lemmon, M., Smith, C. L., Martínez, G., de Vicente-Retortillo, Á., Newman, C. E., et al. (2019). Mars Science Laboratory observations of the 2018/Mars year 34 global dust storm. Geophy. Res. Lett., 46, 71–79.

Haberle, R. M., Leovy, C. B., Pollack, J. B. (1982), Some effects of global dust storms on the atmospheric circulation of Mars, Icarus, 50, 2-3, p. 322-367.

Haberle, R. M., Houben, H. C., Hertenstein, R., Herdtle, T. (1993), A boundary-layer model for Mars - Comparison with Viking lander and entry data, Journal of the Atmospheric Sciences, 50-11, p.1544-1559.

Haberle, R.M., H. Houben, J.R. Barnes, and R.E. Young (1997), A simplified three-dimensional model for Martian climate studis, J. Geophys. Res., 102, 9051-9067.

Haberle, R.M., M.M. Joshi, J.R. Murphy, J.R. Barnes, J.T. Schofield, G. Wilson, M. Lopez-Valverde, J.L. Hollingsworth, A.F.C. Bridger, and J. Schaeffer (1999), General circulation model simulations of the Mars Pathfinder atmospheric structure investigation/meteorology data, J. Geophys. Res., 104, 8957-8974.

Haberle, R. M., Kahre, M. A., Hollingsworth, J. L., Montmessin, F., Wilson, R. J., Urata, R. A., Brecht, A. S., Wolff, M. J., Kling, A. M., Schaeffer, J. R. (2019), Documentation of the NASA/Ames Legacy Mars Global Climate Model: Simulations of the present seasonal water cycle, Icarus, 333, p. 130-164.

Heavens, N. G., Kleinböhl, A., Chaffin, M. S., Halekas, J. S., Kass, D. M., Hayne, P. O., McCleese, D. J., Piqueux, S., Shirley, J. H., Schofield, J. T. (2018), Hydrogen escape from Mars enhanced by deep convection in dust storms, Nature Astronomy, Volume 2, p. 126-132.

Hourdin, F., Forget, F., Talagrand, O. (1995), The sensitivity of the Martian surface pressure and atmospheric mass budget to various parameters: A comparison between numerical simulations and Viking observations, Journal of Geophysical Research, Volume 100, Issue E3, p. 5501-5524.

Kass, D.M., A. Kleinböhl, D.J. McCleese, J.T. Schofield, and M.D. Smith (2016), Interannual similarity in the Martian atmosphere during the dust storm season, Geophys. Res. Lett., 43, 6111-6118.

Kahre, M. A., Murphy, J. R., Haberle, R. M., Montmessin, F., Schaeffer, J. (2005), Simulating the Martian dust cycle with a finite surface dust reservoir, Geophysical Research Letters, Volume 32, Issue 20.

Kahre, M. A., Murphy, J. R., Haberle, R. M. (2006), Modeling the Martian dust cycle and surface dust reservoirs with the NASA Ames general circulation model, Journal of Geophysical Research, Volume 111, Issue E6.

Kahre, M.A., R.J. Wilson, R.M. Haberle, and J.L. Hollingsworth, An inverse approach to modeling the dust cycle with two Mars general circulation models, Mars Dust Cycle workshop, Ames Research Center, CA, September, 2009.
http://humbabe.arc.nasa.gov/MarsDustWorkshop/DustHome.html.

Kahre, M. A., Murphy, J. R., Newman, C. E., Wilson, R. J., Cantor, B. A., Lemmon, M. T. and Wolff, M. J. (2017), The Mars Dust Cycle, The atmosphere and climate of Mars, Edited by R.M. Haberle et al. ISBN: 9781139060172. Cambridge University Press, 2017, p. 229-294

Kahre, M. A., Wilson, R. J., Hollingsworth, J. L., Haberle, R. M., Brecht, A. S., Urata, R. A., Bertrand, T., Kling, A., Jha, V., Batterson, C. M., Steakley, K. (2018), High Resolution Modeling of the Dust and Water Cycles with the NASA Ames Mars Global Climate Model, American Geophysical Union, Fall Meeting 2018.

Kleinböhl, A., Friedson, A. J., Schofield, J. T. (2017), Two-dimensional radiative transfer for the retrieval of limb emission measurements in the martian atmosphere, Journal of Quantitative Spectroscopy and Radiative Transfer, Volume 187, p. 511-522.

Koster, R., Jouzel, J., Suozzo, R., Russell, G., Broecker, W., Rind, D., Eagleson, P. (1986), Global sources of local precipitation as determined by the Nasa/Giss GCM, Geophysical Research Letters, Volume 13, Issue 2, p. 121-124.

Lewis, S. R. and Read, P. L. (2003), Equatorial jets in the dusty Martian atmosphere, J. Geophys. Res. (Planets), 108, E4, 5034.

Madeleine, J.-B., F. Forget, E. Millour, T. Navarro, and A. Spiga (2012), The influence of radiatively active water ice clouds on the Martian climate, Geophys. Res. Lett., 39, L23202, doi10.1092/2012GL053564.

Malin, M. C., Cantor, B. A., & Britton, A. W. (2018a). MRO MARCI weather report for the week of 21 May 2018–27 May 2018, Malin Space Science Systems Captioned Image Release, MSSS-532. Retrieved from http://www.msss.com/msss_images/2018/05/30/
Google ScholarFulltext@NASA ARC Library
Malin, M. C., Cantor, B. A., & Britton, A. W. ( 2018b). MRO MARCI weather report for the week of 28 May 2018–3 June 2018, Malin Space Science Systems Captioned Image Release, MSSS-533. Retrieved from http://www.msss.com/msss_images/2018/06/06/
Google ScholarFulltext@NASA ARC Library
Malin, M. C., Cantor, B. A., & Britton, A. W. ( 2018c). MRO MARCI weather report for the week of 4 June 2018–10 June 2018, Malin Space Science Systems Captioned Image Release, MSSS-534. Retrieved from http://www.msss.com/msss_images/2018/06/13/
Google ScholarFulltext@NASA ARC Library
Malin, M. C., Cantor, B. A., & Britton, A. W. ( 2018d). MRO MARCI weather report for the week of 11 June 2018–17 June 2018, Malin Space Science Systems Captioned Image Release, MSSS-535. Retrieved from http://www.msss.com/msss_images/2018/06/20/
Google ScholarFulltext@NASA ARC Library
Malin, M. C., Cantor, B. A., & Britton, A. W. ( 2018e). MRO MARCI weather report for the week of 18 June 2018–24 June 2018, Malin Space Science Systems Captioned Image Release, MSSS-536. Retrieved from http://www.msss.com/msss_images/2018/06/27/
Google ScholarFulltext@NASA ARC Library
Malin, M. C., Cantor, B. A., & Britton, A. W. ( 2018f). MRO MARCI weather report for the week of 2 July 2018–8 July 2018, Malin Space Science Systems Captioned Image Release, MSSS-538. Retrieved from http://www.msss.com/msss_images/2018/07/11/
Google ScholarFulltext@NASA ARC Library
Malin, M. C., Cantor, B. A., & Britton, A. W. ( 2018g). MRO MARCI weather report for the week of 16 July 2018–22 July 2018, Malin Space Science Systems Captioned Image Release, MSSS-540. Retrieved from http://www.msss.com/msss_images/2018/07/25/
Google ScholarFulltext@NASA ARC Library
Malin, M.C., Cantor, B.A., & Britton, A.W. ( 2018h). MCO MARCI weather report for the week of 3 September 2018–9 September 2018, Malin Space Science Systems Captioned Image Release, MSSS-547. Retrieved from http://www.msss.com/msss_images/2018/09/12/
Google ScholarFulltext@NASA ARC Library

Montabone, L., F. Forget, E. Millour, R.J. Wilson, S.R. Lewis, D. Kass, A. Kleinböhl, M.T. Lemmon, M.D. Smith, and M.J. Wolff (2015), Eight-year Climatology of Dust on Mars, Icarus, http://doi.dx.doi.org/10.1016/j.icarus.2014.12.034.

Montabone 2019, this special issue.

Montmessin, F., Rannou, P., Cabane, M. (2002), New insights into Martian dust distribution and water-ice cloud microphysics, Journal of Geophysical Research (Planets), Volume 107, Issue E6.

Montmessin, F., Forget, F., Rannou, P., Cabane, M., Haberle, R. M. (2004), Origin and role of water ice clouds in the Martian water cycle as inferred from a general circulation model, Journal of Geophysical Research, Volume 109, Issue E10.

Navarro, T., Madeleine, J.-B., Forget, F., Spiga, A., Millour, E., Montmessin, F. and Määtänen, A. (2014), Global climate modeling of the martian water cycle with improved microphysics and radiatively active water ice clouds, J. Geophys. Res.

Nelli, S. M., Murphy, J. R., Feldman, W. C., Schaeffer, J. R. (2009), Characterization of the nighttime low-latitude water ice deposits in the NASA Ames Mars General Circulation Model 2.1 under present-day atmospheric conditions, Journal of Geophysical Research, Volume 114, Issue E11.

Newman, C. E., Lewis, S. R., Read, P. L., Forget, F. (2002), Modeling the Martian dust cycle 2. Multiannual radiatively active dust transport simulations, Journal of Geophysical Research (Planets), Volume 107, Issue E12.

Newman, C. E., Richardson, M. I. (2015), The impact of surface dust source exhaustion on the martian dust cycle, dust storms and interannual variability, as simulated by the MarsWRF General Circulation Model, Icarus, Volume 257, p. 47-87.

Richardson, M. I., Wilson, R. J. (2002), Investigation of the nature and stability of the Martian seasonal water cycle with a general circulation model, Journal of Geophysical Research (Planets), Volume 107, Issue E5.

Sanchez-Lavega, A., Rio-Gaztelurrutia, T., Bernal, J.H. and Delcroix, M. (2019), The onset and growth of the 2018 Martian global dust storm, Geo. Res. Lett.

Savijarvi, H. (1995), Mars boundary layer modeling: Diurnal moisture cycle and soil properties at the Viking Lander 1 Site, Icarus, Volume 117, Issue 1, p. 120-127

Smith, M. D. (2004), Interannual variability in TES atmospheric observations of Mars during 1999-2003, Icarus, Volume 167, Issue 1, p. 148-165.

Smith, M. D., Daerden, F., Neary, L., Khayat, A. (2018), The climatology of carbon monoxide and water vapor on Mars as observed by CRISM and modeled by the GEM-Mars general circulation model, Icarus, Volume 301, p. 117-131.

Spiga, A., Faure, J., Madeleine, J.-B., Määttänen, A., Forget, F. (2013), Rocket dust storms and detached dust layers in the Martian atmosphere, Journal of Geophysical Research: Planets, Volume 118, Issue 4, pp. 746-767

Strausberg, M. J., Wang, H., Richardson, M. I., Ewald, S. P., Toigo, A. D. (2005), Observations of the initiation and evolution of the 2001 Mars global dust storm, Journal of Geophysical Research, Volume 110, Issue E2.

Szwast, M. A., Richardson, M. I., Vasavada, A. R. (2006), Surface dust redistribution on Mars as observed by the Mars Global Surveyor and Viking orbiters, Journal of Geophysical Research, Volume 111, Issue E11.

Toigo, A. D., Richardson, M. I., Wilson, R. J., Wang, H., Ingersoll, A. P. (2002), A first look at dust lifting and dust storms near the south pole of Mars with a mesoscale model, Journal of Geophysical Research (Planets), Volume 107, Issue E7.

Toon, O. B., McKay, C. P., Ackerman, T. P., and Santhanam, K. (1989), Rapid calculation of radiative heating rates and photodissociation rates in inhomogeneous multiple scattering atmospheres. Journal of Geophysical Research, 94 :16287–16301. (cf p. 35, 36 et 45)

Trokhimovskiy, A., Fedorova, A., Korablev, O., Montmessin, F., Bertaux, J.-L., Rodin, A., Smith, M. D. (2015), Mars' water vapor mapping by the SPICAM IR spectrometer: Five martian years of observations, Icarus, Volume 251, p. 50-64.

Vandaele, A. C., et al. (2019), Martian dust storm impact on atmospheric H2O and D/H observed by ExoMars Trace Gas Orbiter, Nature, Volume 568, Issue 7753, p.521-525.

Vincendon, M., Audouard, J., Altieri, F., Ody, A. (2015), Mars Express measurements of surface albedo changes over 2004-2010, Icarus, Volume 251, p. 145-163.

Wang, H., Richardson, M. I., Wilson, R. J., Ingersoll, A. P., Toigo, A. D., Zurek, R. W. (2003), Cyclones, tides, and the origin of a cross-equatorial dust storm on Mars, Geophysical Research Letters, Volume 30, Issue 9.

Wang, H. and Richardson, M. I. (2015), The origin, evolution, and trajectory of large dust storms on Mars during Mars years 24-30 (1999-2011), Icarus, Volume 251, p. 112-127.

Warren, S. G. (1984), Optical constants of ice from the ultraviolet to the microwave, Applied optics, 23(8), 1206-1225.

Wilson, J. R. (1997), Dust transport in the martian atmosphere as simulated by a general circulation model

Advances in Space Research, Volume 19, Issue 8, p. 1290-1290.

Wilson, R.J., and K.P. Hamilton (1996), Comprehensive model simulation of thermal tides in the martian atmosphere, J. Atmos. Sci. 53, 1290-1326.

Wilson, R.J., and M.I. Richardson, 2000: The Martian Atmosphere During the Viking Mission, 1: Infrared Measurements of Atmospheric Temperatures Revisited. Icarus, 145, 555-579.

Wilson, R.J., The role of thermal tides in the Martian dust cycle, European Planetary Science Congress, Madrid, Spain, September, 2012.
http://meetingorganizer.copernicus.org/EPSC2012/EPSC2012-798-1.pdf

Wilson, R.J. Martian dust storms, thermal tides and the Hadley circulation, Abstract 8069, Comparative Climatology of Terrestrial Planets, Boulder, CO, June 2012.
http://www.lpi.usra.edu/meetings/climatology2012/pdf/8069.pdf

Wolff, M. J., Clancy, R. T., Goguen, J. D., Malin, M. C., Cantor, Bruce A. (2010), Ultraviolet dust aerosol properties as observed by MARCI, Icarus, Volume 208, Issue 1, p. 143-155.